\documentclass[%
 aip,
 jmp,%
 amsmath,amssymb,
 reprint,%
]{revtex4-2}

\usepackage{graphicx}
\usepackage{dcolumn}
\usepackage{bm}
\usepackage{amssymb} 
\usepackage{mathtools} 

\begin{document}

\preprint{APS/123-QED}

\title[Autonomous Learning with Chemical Reaction Networks]{Autonomous Learning of Generative Models with Chemical Reaction Network Ensembles}

\author{William Poole}
 \altaffiliation[Currently at ]{Altos Labs, Redwood City, California, USA.}
 \email{wpoole@altoslabs.com}
\affiliation{{Computation and Neural Systems, California Institute of Technology, Pasadena, California, USA}}

\author{Thomas E. Ouldridge}%
\affiliation{ 
Bioengineering, Imperial College London, London, England
}%

\author{Manoj Gopalkrishnan}
\affiliation{%
Electrical Engineering, India Institute of Technology Bombay, Mumbai, India
}%

\date{\today}

\begin{abstract}
Can a micron sized sack of interacting molecules autonomously learn an internal model of a complex and fluctuating environment? We draw insights from control theory, machine learning theory, chemical reaction network theory, and statistical physics to develop a general architecture whereby a broad class of chemical systems can autonomously learn complex distributions. Our construction takes the form of a chemical implementation of machine learning's optimization workhorse: gradient descent on the relative entropy cost function. We show how this method can be applied to optimize any detailed balanced chemical reaction network and that the construction is capable of using hidden units to learn complex distributions. This result is then recast as a form of integral feedback control. Finally, due to our use of an explicit physical model of learning, we are able to derive thermodynamic costs and trade-offs associated to this process.
\end{abstract}

\maketitle

\section{Introduction}
Living organisms demonstrate a remarkable ability to thrive in diverse conditions, survive perturbations, and generally adapt to their environments \cite{kitano2004biological}. In contrast, many synthetic \textit{in vitro} and \textit{in vivo} biochemical circuits require fine-tuning to operate well \cite{del2018future}. These observations are partially reconciled by the substantial experimental and theoretical evidence that biological circuitry has evolved to be inherently adaptive and robust \cite{daniels2008sloppiness,khammash2016engineering}.

The ideas of adaptation and robustness are central tenants in biology and medicine, often discussed under the moniker of homeostasis \cite{cannon1929organization}. The principle of homeostasis states that aspects of the internal environment of an organism (such as a cell) must be held relatively constant in order to maintain the vitality of the organism; in this sense, homeostasis is a dynamic process which enables life to thrive in diverse fluctuating environments \cite{davies2016adaptive}. Homeostasis has frequently been studied mathematically using tools from control theory that emphasize how feedback mechanisms are capable of dynamically adjusting internal variables of a system \cite{stelling2004robustness}. Robust perfect adaptation is a particularly well studied kind of homeostasis by which an internal variable can be held at an exact value despite uncertainty and variability in the the dynamics of the entire system \cite{ferrell2016perfect}; it has been shown formally that for some classes of systems, a sufficient condition for robust perfect adaptation is integral feedback \cite{khammash2021perfect}. Furthermore, robust perfect adaptation has been studied and observed in many biological systems including bacterial chemotaxis, calcium homeostasis, and others \cite{khammash2021perfect,barkai1997robustness,el2002calcium}.The internal model principle of control theory offers another valuable lens for understanding how biochemical systems may be capable of homeostasis. This principle states that in order for a system to adapt to a fluctuating environment, the system must have an internal model of the kinds of fluctuations present in that environment and how they effect the system \cite{francis1976internal,andrews2008approximate,huang2018internal}. 

Machine learning and neural networks, although originally developed for understanding intelligent behavior in biological brains and for engineering analogous behavior in electronic computers, have increasingly been applied as a theoretical framework for complex adaptive behavior in non-neural physical systems \cite{stern2023learning} and, of particular relevance here, in single-cell organisms \cite{fernando2009molecular,dussutour2021learning,gunawardena2022learning}.  Beyond the relevance to biological systems, neural network architectures have inspired attempts to theoretically formulate and experimentally implement artificial chemical systems with similar computational abilities \cite{anderson2021reaction,nagipogu2023survey}.  As a proof-of-principle example, theoretical chemical systems have been described that implement supervised learning of deterministic functions by backpropagation \cite{lakin2023design}. In the context of biological control and homeostasis, it seems has been suggested that learning architectures could be used by chemical systems in order to increase their homeostatic capacity via adaptive internal models\cite{gunawardena2022learning}.

As an alternative to deterministic function evaluation, generative machine learning architectures have emerged as a powerful framework for representing the distributions underlying diverse complex systems \cite{harshvardhan2020comprehensive}. These systems have the abilities to both represent diverse distributions and learn these representations from data \cite{jordan2004introduction}. The utility of generative models to living organisms has been recognized as biologically relevant in neuroscience. Theories such as predictive coding \cite{kilner2007predictive} and the free energy principle \cite{friston2010free} posit that the brain is, at some level, a generative model making predictions about its environment and learning by comparing those predictions to sensory observations. Although some of these principles have begun to be explored in the context of molecular and developmental biology \cite{kuchling2020morphogenesis}, it remains an open question to what extent these theories apply to molecular biology.

Chemical reaction networks (CRNs) provide a powerful formalism rooted in statistical physics to understand biomolecular systems \cite{gunawardena2003chemical}. CRNs have been used to study systems such as gene regulation, signaling, and stochasticity in biology \cite{alon2019introduction,kaern2005stochasticity}. In the context of control, CRNs have been used to study integral feedback in biochemical systems and used to design synthetic biochemical control circuits \cite{del2015biomolecular}. Additionally, when studied as formal models of computation, CRNs have been proven to be Turing universal \cite{soloveichik2008computation} and capable of implementing neural architectures both in theory \cite{anderson2021reaction} and in laboratory settings \cite{qian2011neural}. Importantly, CRNs have been proven to be capable of producing a arbitrary distributions \cite{cappelletti2020stochastic} including Boltzmann machines, an important example of generative models \cite{poole2017chemical,ackley1985learning}. These demonstrations indicate that CRNs have the potential to be powerful generative models; however it is less clear how these generative models can adapt and learn from an external environment.

There is currently limited understanding of learning in stochastic CRNs that represent generative models.  Within a CRN that generates a probabilistic choice based on a single probability, a technique based on operant conditioning effectively adapts the counts of ``weight species'' to optimize a feedback signal \cite{arredondo2022operant}.  We aim at two improvements over this scenario.  First, we are interested learning architectures suitable for classes of CRNs that are capable of representing arbitrarily complex distributions, and second, we wish to exploit stochasticity that arises at equilibrium so as to emphasize the potential energy efficiency of generative models within chemical systems.

In prior related work, we examined how a class of chemical systems called detailed balanced CRNs (which will be formally defined in the Sections \ref{sec:CRNs} and \ref{sec:dbCRNs}) can be viewed as a kind of generative model capable of probabilistic inference \cite{poole2022detailed}. In this paper, we extend these results to show how detailed balanced CRNs are capable of \textit{learning} from a noisy environment when connected to a chemical module that implements the well-known machine learning optimization algorithm gradient descent \cite{ackley1985learning} which we rigorously show is related to feedback control. We then use these results to design a fully autonomous chemical system that is capable of learning complex environmental distributions. Finally, due to the physical nature of this construction, we are able to analyze its thermodynamic costs, adding to a growing body of literature addressing the fundamental physics of machine learning \cite{hjelmfelt1992chemical,goldt2017stochastic,salazar2017nonequilibrium,wolpert2019stochastic}.

\section{Background}
\subsection{Learning}
The \textit{learning} in machine learning is a process by which a mathematical function representing some kind of computation is automatically optimized from data. In the context of this paper, we are interested in a class of functions called graphical generative models, which can be sampled to produce probability distributions $\mathbb{P}(x \mid \theta)$ where $x$ are state variables and $\theta$ are parameters of the distribution \cite{jordan2004introduction}.

Typically, the learning process is formalized to optimize a loss function $\mathcal{L}$ and with an algorithm such as gradient descent in order to find parameters $\theta^*$ from data $\psi$:
\begin{equation}
    \theta^* = \underset{\theta}{\textrm{argmin}} \, \mathcal{L}(\mathbb{P}(x \mid \theta), \psi).
\end{equation}
It is worth emphasizing that learning is a dynamic process in which the parameters $\theta$ are updated iteratively. On the other hand, we are generally not concerned with the dynamics of the underlying generative model and instead will assume that $\mathbb{P}(x \mid \theta)$ is a steady state distribution that can be approximated via sampling.

\subsection{Boltzmann Machines}
\label{sec:BMs}
As an archetypal learning system, we will consider the generative graphical models called Boltzmann Machines (BMs), which have been widely studied \cite{ackley1985learning}. Briefly, a BM is a stochastic neural network with an equilibrium distribution:
\begin{equation}
    \mathbb{P}(x) = \frac{1}{Z} e^{-E(x)}  \quad \quad Z = \sum_{x \in \{0, 1\}^N} e^{- E(x)} \quad \quad E(x) = \sum_{i>j} w_{ij} x_i x_j - \sum_i \theta_i x_i.
\end{equation}
In their simplest form, a BM's nodes are binary, $x_i \in \{0, 1\}$, and the parameters are real-valued $\theta_i, w_{ij} \in \mathbb{R}$. BMs are capable of probabilistic inference (e.g. computing conditional distributions) by holding a subset of their nodes constant. 

Consider the nodes $x$ divided into two groups $u$ and $v$, such that $(u, v) = x$. The conditional distribution $\mathbb{P}(u \mid v)$ can be computed by sampling the BM distribution with Markov Chain Monte Carlo while not allowing $v$ to vary; the variables $v$ are said to be \textit{clamped}.  In contexts where the same variables are being repeatedly clamped, perhaps to different environmental signals at different times, we say that $v$ are the \textit{visible} nodes while $u$ are the \textit{hidden} nodes.
\\\\
The parameters $w$ and $\theta$ of a BM can also be learned via gradient descent on the relative entropy cost function: 
\begin{align}
\label{eq:relative_entropy}
    \mathcal{L}(\mathbb{P}, \psi) = \mathbb{D}(\psi(v) \mid \mid \mathbb{P}(v)) = \sum_{v} \psi(v) \log \frac{\psi(v)}{\mathbb{P}(v)}
\end{align}
where $\mathbb{P}(v) = \sum_u \mathbb{P}(u, v)$ is the marginal of $\mathbb{P}(u, v)$ and the data is distributed according to $\psi(v)$. Note that the relative entropy is not symmetric: $\mathbb{D}(\psi \mid \mid \mathbb{P}) \neq \mathbb{D}(\mathbb{P} \mid \mid \psi)$. In the form of Equation (\ref{eq:relative_entropy}), we can consider $\psi$ to be the ``true'' distribution, while $\mathbb{P}$ is an approximation; this corresponds to the relative entropy being an average over $\psi$: $\mathbb{D}(\psi \mid \mathbb{P}) = \langle \log \frac{1}{\mathbb{P}} \rangle_\psi - \langle \log \frac{1}{\psi} \rangle_\psi$. This equation can be interpreted as  the average excess information required to approximate $\psi$ with $\mathbb{P}$. Hidden units can be explicitly optimized via gradient descent on the relative entropy between the \textit{clamped} distribution $\overline{\mathbb{P}} = \mathbb{P}(u \mid v) \psi(v)$ of the BM held to samples from $\psi$ and the \textit{free} distribution $\mathbb{P}(u, v)$ of the BM:
\begin{equation}
\label{eq:dkl}
    \mathbb{D}( \overline{\mathbb{P}} \mid \mid \mathbb{P}) = \sum_{v, u} \overline{\mathbb{P}}(u, v) \log \frac{\overline{\mathbb{P}}(u, v)}{\mathbb{P}(v, u)} =  \quad \textrm{where} \quad \overline{\mathbb{P}}(u, v) = \mathbb{P}(u \mid v) \psi(v)
\end{equation}
Taking the gradient with respect to $w_{ij}$ and $\theta_i$ results in the update rule which looks identical for both hidden and visible units:
\begin{align}
    \label{eq:learning_rule}
    &\frac{\textrm{d} w_{ij}}{\textrm{d} t} = \frac{\partial \mathbb{D}( \overline{\mathbb{P}} \mid \mid \mathbb{P})}{\partial w_{ij}} = \epsilon ( \langle x_i x_j \rangle_{\overline{\mathbb{P}}} - \langle x_i x_j \rangle_{\mathbb{P}}) 
    &&\frac{\textrm{d} \theta_i}{\textrm{d} t} = \frac{\partial \mathbb{D}( \overline{\mathbb{P}} \mid \mid \mathbb{P})}{\partial \theta_i} = \epsilon( \langle x_i \rangle_{\overline{\mathbb{P}}} - \langle x_i \rangle_{\mathbb{P}}) 
\end{align}
Here $\langle \cdot \rangle_\mathbb{P}$ and $\langle \cdot \rangle_{\overline{\mathbb{P}}}$ denote the expected value with respect to the free and clamped distributions, respectively, and $\epsilon$ is the learning rate. Notice that the learning algorithm - gradient descent on the relative entropy - gives rise to dynamics for the parameters $w$ and $\theta$.

\subsection{Chemical Reaction Networks}
\label{sec:CRNs}
Chemical Reaction Networks (CRNs) are a common model of well-mixed chemical environments, meaning that continuous spatial dynamics are neglected \cite{gillespie1992rigorous}. CRNs are a widely used modeling language for synthetic and systems biology and have been studied from many perspectives including computer science, mathematics, and statistical physics \cite{soloveichik2008computation,gunawardena2003chemical,gillespie1992rigorous}. We denote a CRN to be a set of species $S_i \in \mathcal{S}$, reactions $\mathcal{R}$ and rate constants $k$. Reactions convert one multi-set of species to another: $I^r \xrightarrow{k_r} O^r$ where $I^r_i$ and $O^r_i$ are vectors denoting the number of species $S_i$ in the the input and output of the reaction $r$, respectively. In the limit of a system with infinite volume but finite concentrations, a CRN defines deterministic dynamics of the species' concentrations:
\begin{equation}
    \frac{\textrm{d}[S_i]}{\textrm{d}t} = \sum_r M_i^r \eta_r([S]) \quad 
    \textrm{with} \quad \eta_r([S]) = k_r \prod_i [S_i]^{I_i^r}.
\end{equation}

Here, $M^r_i = O^r_i - I^r_i$ is the stoichiometric matrix and we will assume that all reactions occur according to mass action rates $\eta_r([S])$ unless explicitly noted otherwise\footnote{For convenience, we will specify concentrations, volume, and time in units such that a concentration of 1 indicates 1 molecule in a volume of 1, and a bimolecular reaction with both distinct reactants at concentration 1 will take place at rate 1.}.
CRNs can also be considered stochastically in a finite volume $V$ in which case the dynamics of the probability that the species $S$ will have counts $s$ at time $t$ is modeled using the chemical master equation:
\begin{equation}
    \label{eq:cme}
    \frac{\textrm{d}\mathbb{P}(s, t)}{\textrm{d}t} = \sum_r \mathbb{P}(s-M^r, t)\rho_r(s-M^r) - \mathbb{P}(s, t)\rho_r(s) \quad \textrm{with} \quad \rho_r(s) = \frac{k_r}{V^{|I^r|-1}} \prod_i \frac{s_i !}{(s_i-I_i^r)!}.
\end{equation}
Here, all reactions occur with probability proportional to their mass-action propensity $\rho_r(s)$, 

which is equal\footnote{
Physically, rate constants in deterministic and stochastic equations have different meanings (due to being expressed with different units): the deterministic equations are volume-independent, while the values in the stochastic context must depend on volume (and can be related to the deterministic values given a choice of volume).}
to the mass-action rate $\eta_r([S])$ in the limit $V \rightarrow \infty$ while $s_i=V [S_i]$ with constant concentrations.
We will often be interested in the stationary or steady-state distribution $\mathbb{P}^*(s)$ found by equating (\ref{eq:cme}) to 0 or by simulating the CRN using the Gillespie algorithm until convergence \cite{gillespie2007stochastic}. It is worth noting that not all CRNs have unique or well-defined steady-state distributions - a technicality that will play a role in this paper later.

\subsection{Detailed Balanced Chemical Reaction Networks}
\label{sec:dbCRNs}
Detailed balanced CRNs (dbCRN) are a subclass of stochastic CRNs with the following properties:
\begin{itemize}
    \item Each species $S_i \in \mathcal{S}$ has a formal energy $G_i$
    \item All reactions are reversible meaning if $I \xrightarrow{k^+} O \in \mathcal{R}$ then $O \xrightarrow{k^-} I \in \mathcal{R}$.
    \item Reaction rates obey $\frac{k^+}{k^-} = e^{-\Delta G}$, where $\Delta G = \sum_i G_i(O_i- I_i)$.
\end{itemize}
A dbCRN's stationary distribution is an equilibrium distribution when there is no net energy flow or entropy production at steady state. Detailed balanced CRNs may or may not be mass-conserving, meaning their equilibrium distribution can correspond to both the canonical or grand canonical ensembles of statistical physics. Regardless of ensemble, all dbCRNs have a steady-state  distribution with Poisson form written in terms of the free energy function $\mathcal{G}$ ~\cite{anderson2010product}:
\begin{equation}
    \label{eq:product_poisson}
    \pi(s) = \frac{1}{Z}\prod_i \frac{e^{-G_is_i}}{s_i!} = \frac{1}{Z} e^{-\mathcal{G}(s)} \quad \quad Z = \sum_{s \in \Gamma_{s(0)}} e^{-\mathcal{G}(s)} \quad \quad \mathcal{G}(s) = \sum_i \mathcal{G}_i(s) =  \sum_i G_i s_i + \log s_i !
\end{equation}
Here, $Z$ is called the partition function and $\Gamma_{s(0)}$ is called the reachability class that represents the set of all states reachable by any sequence of reactions in $\mathcal{R}$ from the initial condition $s(0)$. In this paper, we will emphasize that there are classes of dbCRNs with the same species and reactions but different rate constants reflecting different species' energies. Let $\mathcal{D} = (\mathcal{S}, \mathcal{R}, k)$ be a detailed balanced CRN, then $\mathcal{D}_G = (\mathcal{S}, \mathcal{R}, k_G)$ is the same set of species and reactions with the rate constants $k$ changed to $k_G$ in order to be compatible with the new energies $G$. We note that $k_G$ is not necessarily unique, as the forward and backward rate constants for any single reaction can always be re-scaled together by the same factor without breaking detailed balance, and thus the re-scaled dbCRN achieves the same equilibrium distribution.

\subsubsection{Detailed Balanced CRNs Can Model Complex Environments}
Despite the canonical product Poisson form, specific dbCRNs are in fact capable of representing complex distributions \cite{poole2017chemical,cappelletti2020stochastic}. In recent work, we unify these observations and provide conditions under which a dbCRN can represent distributions that appear to be different from a product Poisson distribution \cite{poole2022detailed}. Briefly, for a dbCRN to be far from product Poisson requires reactions and an initial condition which toogether restrict the size of the reachability class relative to the full positive integer lattice. This fact is relevant because it provides an existence proof and suggests some design criteria for determining dbCRN reaction networks that may be capable of learning complex and far-from-Poisson distributions.

\subsubsection{Inference with Detailed Balanced CRNs}
\label{subsec:inference_with_dbcrns}
\textit{Any} dbCRN is capable of probabilistic inference. First, we restate some definitions from \cite{poole2022detailed}. Split the species in the dbCRN into disjoint sets of free and clamped species $(\mathcal{S}^F, \mathcal{S}^C) = \mathcal{S}$.
\begin{itemize}
    \item A \textbf{clamped} dbCRN $\mathcal{D}_{G^C} = (\mathcal{S}, \mathcal{R}, k_{G^C})$ is a dbCRN with its energies $G^C = G + \Delta$ where typically $\Delta_i \neq 0$ for any clamped species $S_i \in \mathcal{S}^C$. The equilibrium distribution of this dbCRN is denoted $\pi_C$ and is given by equation (\ref{eq:product_poisson}) with energy vector $G^C$ replacing $G$.
    \item A \textbf{potentiated} dbCRN $\mathcal{D}_{G}^\mathcal{P} = (\mathcal{S} \cup \mathcal{P}, \mathcal{R}^\mathcal{P}, k_G)$ is a dbCRN where each clamped species $S_i \in \mathcal{S}^C$ is coupled to a potential species $P_i \in \mathcal{P}$ such that whenever $S_i$ is created or destroyed by any reaction, its corresponding potential species is created or destroyed with it.  The bulk concentration $[P_i]$ of potential species $P_i$ is held constant by the stochastic system being in contact with an infinite potential bath at equilibrium \cite{schmiedl2007stochastic,polettini2014irreversible}. 
    Therefore, every potentiated dbCRN is detailed balanced with equilibrium distribution $\pi(s \mid [P]) = \pi^P(s)$ which depends on the potential species' concentrations $[P]$. For example, the potentiated reactions of the one-reaction dbCRN with $\mathcal{R} = \{\emptyset \xrightleftharpoons[]{} S_i\}$ are $\mathcal{R}^\mathcal{P} = \{\emptyset \xrightleftharpoons[]{} S_i + P_i\}$.
\end{itemize}
In \cite{poole2022detailed} we showed that the following three notions of clamping are equivalent:
\begin{enumerate}
    \item \textbf{Sample Clamping}: Collecting a biased sample of a dbCRN $\mathcal{D}_G$'s dynamics such that the sampled mean $\langle s^C \rangle = \overline{c}$.
    The bias is achieved by running unbiased trajectories and rejecting those whose means deviate from the target, in the appropriate limit.
    \item \textbf{Energy Clamping}: Varying the energies $\Delta$ in a clamped dbCRN $\mathcal{D}_{G_C}$ such that $\langle s^C \rangle_{\pi_C} = \overline{c}$.
    \item \textbf{Potential Clamping}: Varying the potential species' concentrations $[P_i]$ in a potentiated dbCRN $\mathcal{D}_{G}^\mathcal{P}$: $\Delta_i  = G_i^P + \log [P_i]$ where $G_i^P$ is the energy of the potential species $P_i$.
\end{enumerate}
In other words, conditioning on a dbCRN's clamped species having a particular mean value, modulating the energies of a clamped dbCRN, and changing the concentrations of  the potential species of potentiated dbCRN are all equivalent forms of computing a conditional distribution $\pi(s^F | \langle s^C \rangle = \overline{c})$. These computations all are forms of probabilistic inference well suited for chemical systems and capable of highly complex computations when used on dbCRNs equilibrium distributions with highly restricted reachability classes \cite{poole2022detailed}. 

\subsubsection{In-silico Learning for Detailed Balanced CRNs}
In our previous work on chemical Boltzmann machines~\cite{poole2017chemical}, we showed that a version of the BM learning rule (\ref{eq:learning_rule}) also works for dbCRNs:
\begin{equation}
    \label{eq:dbCRN_learning_rule}
    \frac{\textrm{d} G_i}{\textrm{d} t} 
    =
    \frac{\partial \mathbb{D}(\overline{\pi} \mid \mid \pi)}{\partial G_i} 
    =
    \epsilon (\langle s_i \rangle_{\overline{\pi}} - \langle s_i \rangle_{\pi}).
\end{equation}
Naively, unlike the BM learning rule, equation (\ref{eq:dbCRN_learning_rule}) appears to only utilize the means of each species and not second moments. However, a detailed balanced constructed called the \textit{Edge-Species Chemical Boltzmann Machine} (ECBM) has been shown to also learn second moments. This can be seen by example in a two-node ECBM (which is generalized to 

arbitrary graphs in \cite{poole2017chemical} and is an example of a dbCRN which produces far-from-Poisson distributions):
\begin{align}
    S_1^0 + S_2^0 \xrightleftharpoons[]{} S_1^1 + S_2^0 \quad \quad S_1^1 + S_2^0 + S_W^0 \xrightleftharpoons[]{} S_1^1 + S_2^1 + S_W^1 \\
    S_1^0 + S_2^0 \xrightleftharpoons[]{} S_1^0 + S_2^1 \quad \quad S_1^0 + S_2^1 + S_W^0 \xrightleftharpoons[]{} S_1^1 + S_2^1 + S_W^1.
\end{align}
Here, each node $i \in \{1, 2\}$ of a BM is represented by an \textit{off} species $S_i^0$ and an \textit{on} species, $S_i^1$. The edge species $S_W^0$ and $S_W^1$ similarly has \textit{off} and \textit{on} states and relates to the energy term $w_{12}$ in a BM. The key observation is that this system has an emergent conservation law $s_1^1 s_2^1 = s_W^1$ which occurs when $s_1^0+s_1^1 = s_2^0+s_2^1 = s_W^0 + s_W^1 = 1$ and the CRN starts in a state which respects these laws. This conservation law allows the the energy $G_{S_W^1}$ to be updated based upon the mean of $\langle s_W^1 \rangle$ or, equivalently, based on the the second moment $\langle s_1^1 s_2^1 \rangle$. This directly relates the energy $w_{ij}$ to the species energy $G_{W_{ij}^1}$ of an ECBM and shows that this generative model is capable of representing second-order moments. Further, with this constraint, equation (9) takes the exact form of the classical Boltzmann machine learning rule. 

In previous work the dbCRN learning rule was presented as a formal operation which could be performed \textit{in silico} but 

no explicit construction of a formal CRN that autonomously implemented the learning was given.
In this work, we will develop an autonomous CRN implementation of the learning rule (\ref{eq:dbCRN_learning_rule}) which works on any dbCRN including networks with hidden species. To do this, we will make use of the of the potential clamping construction in the context of many vesicles (or proto-cells) interacting with a large environment, necessitating use of an augmented hybrid CRN model that includes both deterministic and stochastic aspects.

\section{Results Overview}
The potentiated dbCRN implementation provides a mechanism to implement energy clamping and hence inference without having to modify the underlying energies of chemical species (which are, in a certain sense, physical constants). Instead, the concentrations of the potential species can be controlled. We will first derive a non-detailed balanced potential clamping CRN that can tune the potential species $[P_i]$ such that the mean of $S_i$ matches the mean of a target species $Q_i$, which may be thought of as the environment. We examine a fully stochastic implementation of this system and show that it can work in some cases but can also fail catastrophically due to extinction events of the potential species. To remedy this pathological behavior, we consider a novel deterministic-stochastic hybrid model consisting of many stochastic vesicles. This hyprid CRN is shown to exactly implement the learning rule (\ref{eq:learning_rule}) and provides a chemical mechanism by which clamping may occur dynamically. Then, by combining a number of these potential clamping CRNs together with potentiated dbCRN modules and an environment representing data, we show how any potentiated dbCRN can be embedded into a learning CRN architecture that is able to learn potentials for both hidden and visible species in order to approximate an environmental distribution. We emphasize that the learning CRN is an autonomous CRN which automatically learns from the environment, or in other words, that we have rewritten a fundamental machine learning algorithm as a CRN. Finally, we will use this model to provide some simple results on the thermodynamics of learning.

\section{Stochastic Potential Clamping CRN}
First, we consider a potentiated dbCRN, $\mathcal{D}_G^\mathcal{P} = (\mathcal{S}\cup\mathcal{P}, \mathcal{R}_\mathcal{P}, k)$, as defined in the previous section. We will then add the following species and reactions to $\mathcal{D}_G^\mathcal{P}$ and construct a new potential clamping CRN. For each potential species, $P_i \in \mathcal{P}$, add in target species $Q_i \in \mathcal{Q}$ that later will be allowed to fluctuate according to some distribution $\psi(q)$.  We are agnostic about how $\psi$ is generated---it could be equilibrium or non-equilibrium.  Additionally add the non-detailed balanced reactions, $\mathcal{T}^P_{\mathcal{S}, \mathcal{Q}}$, that clamp the species $\mathcal{S}$ to the species $\mathcal{Q}$ using the potentials $\mathcal{P}$:
\begin{align}
\label{eq:learningCRN1}
\mathcal{T}^P_{\mathcal{S}, \mathcal{Q}} =
\left\{
    \begin{matrix}
    P_i + Q_i \xrightarrow{\epsilon k^Q_i} Q_i \\
    P_i + S_i \xrightarrow{\epsilon k^S_i} 2 P_i + S_i
    \end{matrix}
\right \}.
\end{align}
Here, $\epsilon \ll \max(k)$ act as a scaling parameter so the potential clamping reactions occur more slowly than any of the detailed balanced reactions in $\mathcal{R}_\mathcal{P}$ such that sub-network $\mathcal{D}_G^\mathcal{P}$ is at quasi-equilibrium relative to the chemical reaction network $\mathcal{T}^P_{\mathcal{S}, \mathcal{Q}}$. Note that unlike in our previous work, the potential species are no longer assumed to be at a constant concentration. Denote this new CRN $\mathcal{C}_G^\mathcal{Q} = (\mathcal{S}\cup\mathcal{P}\cup\mathcal{Q}, \mathcal{R}_\mathcal{P} \cup \mathcal{T}^P_{\mathcal{S}, \mathcal{Q}}, k)$. 
As defined, this CRN does not change the counts of any target species $Q_i$ as they appear only catalytically; later we will couple $\mathcal{C}_G^\mathcal{Q}$ to another CRN representing the (equilibrium or non-equilibrium) environment that drives the $Q_i$ to fluctuate according to $\psi(q)$ but has no other species in common.  The behavior of $\mathcal{C}_G^\mathcal{Q}$ with fixed counts of $Q_i$ provides insight into the behavior of the coupled system when the environmental fluctuations are even slower than the potential clamping reactions.

Notice that $\mathcal{C}_G^\mathcal{Q}$ is no longer detailed balanced and instead of an equilibrium distribution, it may have a non-equilibrium steady state, $\mathbb{P}(s, q, [P])$ which may be hard to analyze. Indeed, in SI \ref{eq:pot_clamp_example} we provide an example illustrating that this steady state distribution may not be unique. This potentially pathological behavior makes a fully stochastic version of our system intractable as a model of learning and motivates the use of an ensemble of potential clamping CRNs.

\begin{figure}
    \centering
    \includegraphics[width = .95\textwidth]{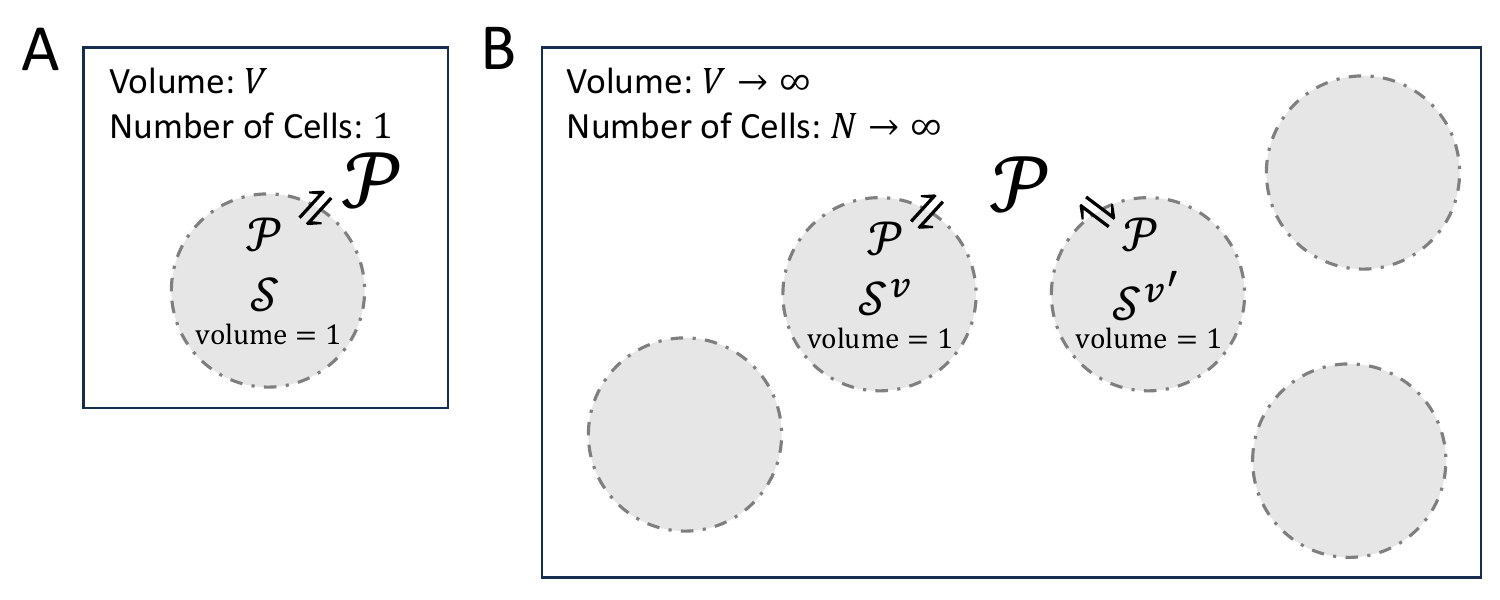}
    \caption{A. A cartoon illustration of a single unit volume vesicle in a large but finite volume $V$ with internal species $\mathcal{S}$ and potential species $\mathcal{P}$. B. A cartoon illustration of the vesicle ensemble where $N\to\infty$ vesicles with distinct internal species $\mathcal{S}^v$ share potential species $\mathcal{P}$ in an infinite volume.}
    \label{fig:vesicle_ensemble_cartoon}
\end{figure}

\section{Ensemble Potential Clamping CRN}
The fundamental reason why the stochastic potential clamping CRN exhibits multiple steady-states behavior is the possibility for extinction (meaning 0 counts) of the $P_i$ species and the $S_i$ species simultaneously resulting in the dynamics of the CRN halting completely due to the fact that $P_i$ and $S_i$ are inputs to every reaction in the system by construction. To remedy this, we will consider the deterministic limit of an ensemble of $N$ identically and independently distributed vesicles (with unit volumes for simplicity) in a large volume $V$ as illustrated in figure (\ref{fig:vesicle_ensemble_cartoon}). Each vesicle $v$ will contain its own copy of the same potentiated dbCRN consisting of species $\mathcal{S}^v$ that cannot diffuse through the vesicle's membrane and reactions $\mathcal{R}_{\mathcal{P}}^v$. However, all vesicles share the same potential species $\mathcal{P}$ that rapidly diffuse through their membranes. Crucially, any species $S_i^v$ and $S_i^{v'}$ are coupled to the same potential species $P_i$. Intuitively, an ensemble of independent vesicles in a large volume will have a very low chance of many simultaneous extinction events. 
As we prove in SI \ref{si:qev_ensemble}, in the large $N$ and large $V$ limit, with the detailed balanced reactions occurring much faster than the potential clamping reactions, and many more counts of $P_i$ than $S_i^v$ for every vesicle $v$, this entire system can be reduced to a non-mass-action CRN governing the potential species:

\begin{equation}
\label{eq:potential_clamping_reactions}
    P_i + Q_i \xrightarrow{\epsilon k_i^Q} Q_i 
    \quad \quad \quad
    P_i + S_i \xrightarrow{\rho_i^S(s, p, \pi^P)} 2 P_i + Si 
    \quad \quad \quad
    \rho_i^S(s, p, \pi^P) = \epsilon k_i^S p_i \nu \langle s_i \rangle_{\pi^P}.
\end{equation}

\begin{figure}
    \centering
    \includegraphics[width = .95\textwidth]{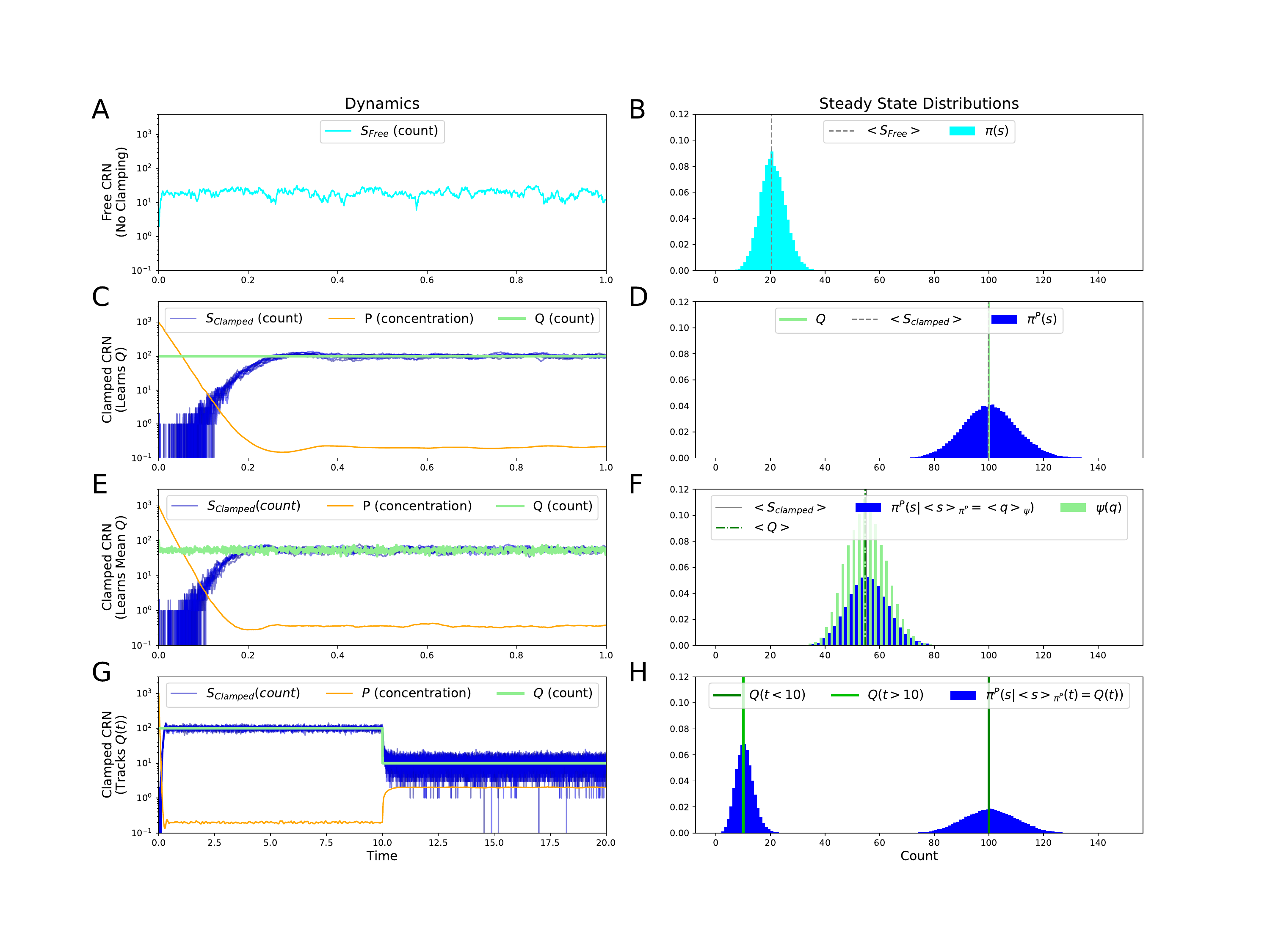}
    \caption{Dynamics and steady state distribution of potential clamping a single species $S$ to $Q$ using potential species $P$. A. The free (unclamped) dynamics of a dbCRN $\emptyset \xrightleftharpoons S$. B. Free steady state distribution of $S$. C. Dynamics of $S$ clamped to a constant value $Q$. D. The mean clamped steady state distribution of $S$ matches $Q$. E. Dynamics $S$ clamped to a quickly varying $Q$. F. The mean of the clamped steady state distribution matches the mean of $Q$. G. Dynamics of $S$ clamped to a slowly varying $Q$ exhibits reference tracking. H. The clamped distribution of $S$ has a mode matching each value of $Q$. Notice that in all these different cases, the potential clamping reactions are very effective at matching $S$ to $Q$.}
    \label{fig:clamping_crn_dynamics}
\end{figure}

Here, the first reaction uses the standard mass-action propensity, $\nu = N/V$ is the vesicle concentration, and $\langle s_i \rangle_{\pi^P}$ is the expected value of $S_i$ with respect to the quasi-equilibrium distribution $\pi^P$. Due to the inclusion of expected values, it is unclear how to simulate this CRN stochastically.
However, as we show in SI \ref{si:qev_ensemble}, this CRN can be interpreted as a hybrid deterministic-stochastic system in the limit: $\epsilon \to 0$, $V\to\infty$, $N\to\infty$ and $p_i \to \infty$ such that $[P_i] = \frac{p_i}{V}$ and $\nu = \frac{N}{V}$. This limit results in the ODEs:
\begin{align}
    \frac{\textrm{d} [P_i]}{\textrm{d}t} =  \epsilon \left( k_i^S [P_i] \nu \langle s_i \rangle_{\pi^P} - k_i^Q [P_i][Q_i] \right).
\end{align}
This ODE is equivalent to the moment learning rule as seen by rewriting equation (\ref{eq:dbCRN_learning_rule}) using the explicit form of the free energy change $\Delta_i = G^P_i + \log P_i$:
\begin{align}
    \frac{\textrm{d}}{\textrm{d}t} \Delta_i = \frac{\textrm{d}}{\textrm{d}t} (G^P_i + \log P_i) = \frac{1}{P_i} \frac{\textrm{d} [P_i]}{\textrm{d}t}  = \epsilon \left( k_i^S \nu \langle s_i \rangle_{\pi^P} - k_i^Q [Q_i] \right).
\end{align}
The constant $\epsilon$ is analogous to a learning rate. The extra constants $k_i^S$, $k_i^Q$ and $\nu$ can be interpreted as scale factors controlling the relative concentrations of the target species $Q$, the concentrations of vesicles $\nu$, and the counts of $S$. In other words, at steady state $S_i$ is clamped to a mean value proportional to $Q_i$ or $\langle Q_i \rangle$ as illustrated in figure \ref{fig:clamping_crn_dynamics}:
\begin{align}
\label{eq:learning_crn_ss}
    \frac{\textrm{d} [P_i]_{ss}}{\textrm{d}t} = 0 \implies 
    \langle s_i \rangle_{\pi^P} = \left\{
    \begin{matrix} 
    \alpha_i [Q_i] & Q_i \textrm{ is constant}
    \\
    \alpha_i \langle Q_i \rangle & Q_i \textrm{ varies quickly}
    \end{matrix}
    \right\}
    \quad \textrm{with} \quad \alpha_i = \frac{k_i^Q}{k_i^S \nu}
\end{align}
The dynamics of this system can also be solved assuming $[Q]$ is constant:
\begin{align}
    \label{eq:clamping_dynamics}
    \frac{\textrm{d}[P_i](t)}{[P_i](t)} = \epsilon (k_i^S \nu \langle s_i \rangle_{\pi^P} - k_i^Q \langle q_i(t) \rangle_{\psi}) \textrm{d}t \implies
    [P_i](t) = C e^{\epsilon (k_i^S \nu \langle s_i \rangle_{\pi^P} - k_i^Q  [Q_i](t))}
\end{align}
where $C$ is a constant of integration. In words, the CRN $\mathcal{C}_G^\mathcal{Q}$ will vary $[P_i]$, which is equivalent to varying the energy of $S_i$ via the chemical potential $\mu_i$: $\Delta_i = \mu_i = G_i^P + \log [P_i]$. Furthermore, if the rates $\epsilon$ are small enough and the correct initial condition is chosen, this system may reach a fixed point $[P_i]_{ss}$ such that equation (\ref{eq:learning_crn_ss}) is satisfied---meaning that the mean of $S_i$ is equal to the mean of $Q_i$ multiplied by the scaling factor $\alpha_i$. The dynamics also illustrate that clamping will be fast with exponential convergence to the correct value provided there is no error in estimating the expected value $\langle s \rangle_{\pi^P}$. However, as these distributions become noisier (meaning not enough time scale separation), the exponential will amplify fluctuations and learning becomes a biased random walk. Importantly, some level of fluctuations in the dynamics may actually help convergence to an optimal solution, just as stochastic gradient descent using mini-batches frequently converges better than deterministic gradient descent implementations by avoiding spurious local minima. Figure \ref{fig:clamping_crn_dynamics} illustrates the dynamics of the potential clamping reactions when applied to a birth death process of $S$ which is shown un-clamped in panel A. In panel C, the target species $Q$ is held constant and $S$ is clamped to its value $[Q]$. In panel E, $Q$ fluctuates quickly and $S$ is clamped to $\langle Q \rangle_\psi$. Finally, in panel G, the $Q$ varies slowly and the potential clamping CRN can be viewed as a form of integral feedback control which causes $S$ to track the reference signal $[Q_i](t)$ as discussed in more detail below.

Intuitively, it can be seen that the potential clamping CRN is performing feedback control. Let $\alpha_i = k_i^Q / k_i^S$. When $\alpha_i Q_i > S_i$, the net production rate of $P_i$ will be negative. Having fewer $P_i$ molecules in the system will favor reactions which produce $S_i$. On the other hand, when $S_i > \alpha_i Q_i$, $P_i$ will be produced, favoring reactions which consume $S_i$. Figure \ref{fig:clamping_crn_dynamics} shows the behavior of the potential clamping reactions coupled to a birth death process $\emptyset \xrightleftharpoons[]{} S + P$ and shows that this model is capable of reference tracking. Furthermore, in SI \ref{SI:feedback_control} we formally prove the potential clamping CRN is similar to feedback control by showing that the derivative of the expected value of $S_i$ proportional to the error.

\section{Autonomous Learning CRNs with Hidden Units}
\begin{figure}
    \centering
    \includegraphics[width = .9\textwidth]{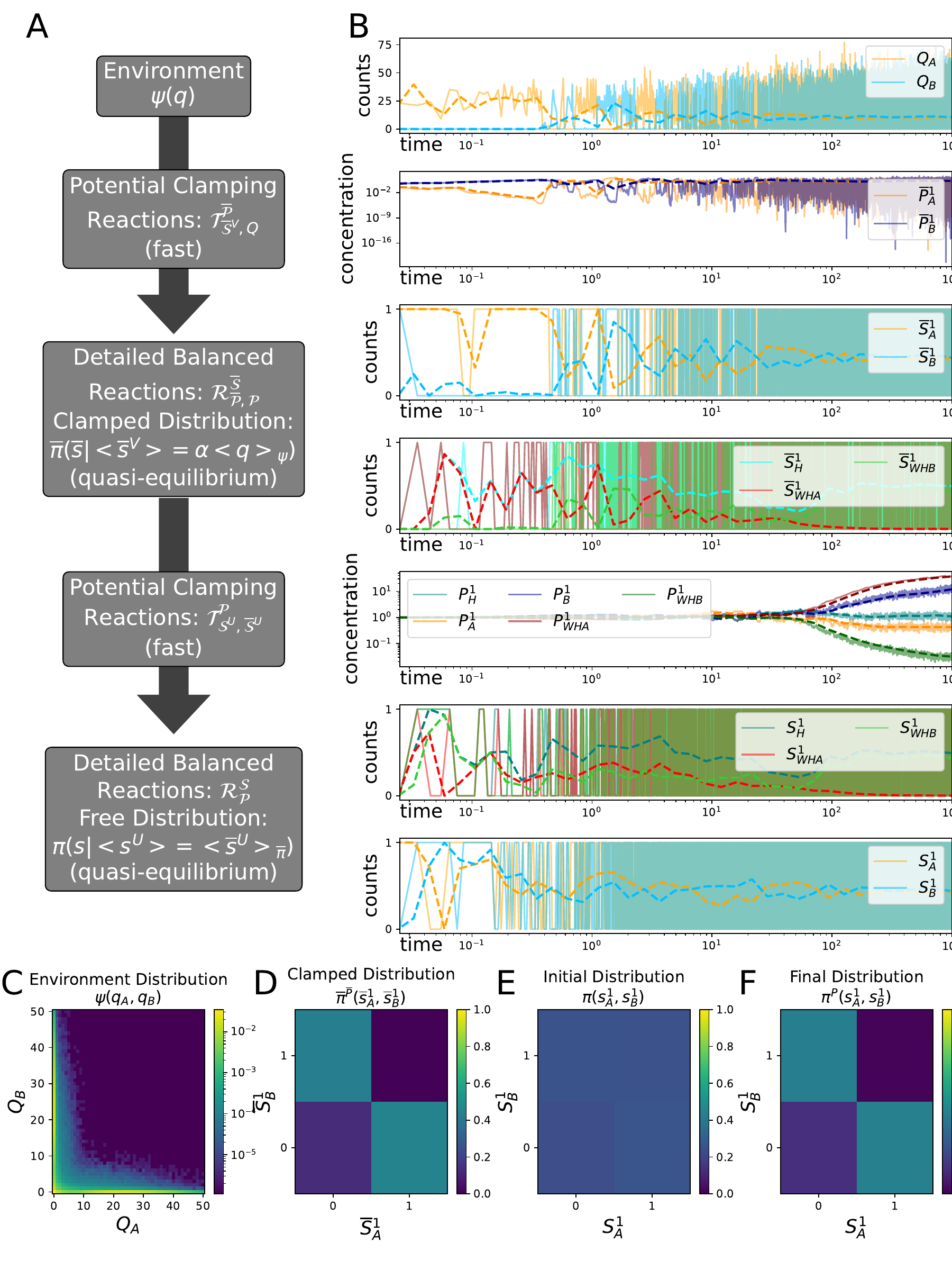}
    \caption{A. The learning architecture coupling dbCRNs and the environment together with potential clamping reactions. 
    Arrows point from the target species to the clamped species. B. Trajectories from this architecture applied to learn an XOR distribution with a 3-node chemical Boltzmann machine from a bistable genetic toggle switch. Note that the X-axis (time) is logarithmic. Solid lines show trajectories for a single training iteration. Dashed lines denote the time-binned means of similarly colored species. C. The steady state distribution of the bistable toggle switch. D. The distribution obtained from clamping to the environment. E. The initial distribution of the chemical Boltzmann machine before training. F. The final distribution of the chemical Boltzmann machine after training.}
    \label{fig:learning_crn}
\end{figure}

In this section, we show how to create an autonomous CRN capable of learning by dynamically adjusting potential species so that an internal potentiated dbCRN matches an environmental distribution. To do this, we use potential clamping reactions first to produce a potentiated dbCRN clamped to the environment and then to couple this clamped potentiated dbCRN to a free potentiated dbCRN. We will argue that this construction is an implementation of the learning rule as a continuous-time online process.

The full construction is as follows. Let the environmental distribution $\psi$ have visible species $\mathcal{Q}^V$. Let $\overline{\mathcal{C}}_G^\mathcal{Q} = (\mathcal{Q}  \cup \overline{\mathcal{S}} \cup \overline{\mathcal{P}} \cup \mathcal{P},\mathcal{R}^{\overline{\mathcal{S}}}_{\overline{\mathcal{P}}, \mathcal{P}} \cup \mathcal{T}^{\overline{\mathcal{P}}}_{\overline{\mathcal{S}}, \mathcal{Q}}, k)$ be a clamped potentiated dbCRN with a subset of the its species  $\overline{\mathcal{S}}^V \subseteq \overline{\mathcal{S}}$ clamped to environmental species $\mathcal{Q}^V \sim \psi$ via the potentials $\overline{\mathcal{P}}$. A second set of potentials $\mathcal{P}$ couple $\overline{\mathcal{C}}_G^\mathcal{Q}$ with another clamped potentiated dbCRN $\mathcal{C}_G^{\overline{S}} = (\overline{\mathcal{S}} \cup \mathcal{S} \cup \mathcal{P}, \mathcal{R}^\mathcal{S}_\mathcal{P} \cup \mathcal{T}^\mathcal{P}_{\mathcal{S}, \overline{\mathcal{S}}}, k)$. The species $\mathcal{S}$ are clamped to the values of the species $\overline{\mathcal{S}}$ using the second set of potentials $P$. This construction produces one large learning CRN $\mathcal{L}_G^\mathcal{Q} = (\mathcal{Q} \cup \overline{\mathcal{S}} \cup \mathcal{S} \cup \overline{\mathcal{P}} \cup \mathcal{P}, \mathcal{R}^{\overline{\mathcal{S}}}_{\overline{\mathcal{P}}, \mathcal{P}} \cup \mathcal{R}^\mathcal{S}_\mathcal{P} \cup \mathcal{T}^{\overline{\mathcal{P}}}_{\overline{\mathcal{S}}, \mathcal{Q}} \cup \mathcal{T}^\mathcal{P}_{\mathcal{S}, \overline{\mathcal{S}}}, k)$ illustrated in figure \ref{fig:learning_crn}A. 

Although seemingly complicated, $\mathcal{L}_G^\mathcal{Q}$ is actually implementing a version of the moment learning algorithm of Boltzmann machines (equation \ref{eq:learning_rule}) as a continuous-time online process. We illustrate this with an example in Figure \ref{fig:learning_crn} where a two three-node CBMs (one clamped, one free) use two sets of potential clamping reactions to learn the binary representation of the steady state distribution generated by a stochastic bistable toggle switch~\cite{gardner2000construction}. The environment has two visible species $Q_A$ and $Q_B$ fluctuating according to a stochastic CRN. The CRN version of the Boltzmann machine learning rule requires two coupled 3-node Chemical Boltzmann Machines (CBM). The \textit{clamped} CBM has two visible nodes represented by pairs of binary species $\overline{S}_B^\alpha$ and $\overline{S}_B^\alpha$, and a single hidden node represented by the species $\overline{S}_H^\alpha$. Edges that connect the hidden node to each visible are repreented by the species: $\overline{S}_{W^{HA}}^\alpha$ and $\overline{S}_{W^{HB}}^\alpha$. Here $\alpha \in \{0, 1\}$ are used to denote the off and on forms of each node. The species $\overline{S}_A^1$ and $\overline{S}_B^1$ are clamped to the environmental species $Q_A$ and $Q_B$, respectively, via the potential clamping reactions $\mathcal{T}^{\overline{\mathcal{P}}}_{\overline{\mathcal{S}}, \mathcal{Q}}$ which modulate the species $\overline{P}_A$ and $\overline{P}_B$. These of potential clamping reactions are effectively copying the environment into the clamped CBM. The \textit{free} CBM has the same topology as the \textit{clamped} CBM: nodes are represented by species $S^\alpha_B$, $S^\alpha_B$, and $S^\alpha_H$ with edges $S^\alpha_{W^{HA}}$ and $S^\alpha_{W^{HB}}$. The free CBM is clamped to the clamped CBM via the potential learning reactions $\mathcal{T}^\mathcal{P}_{\mathcal{S}, \overline{\mathcal{S}}}$ which modulate the species $P_A$, $P_B$, $P_H$, $P_{W^{BA}}$, and $P_{W^{HB}}$. This second set of potential clamping reactions is allowing the free CBM to learn from the distribution of the clamped CBM.

Figure \ref{fig:learning_crn}C shows the environmental distribution $\psi$ that produces a bimodal distribution with modes around $Q_A, Q_B = (25, 0)$ and $Q_A, Q_B = (0, 25)$. The species $\overline{\mathcal{S}}^V$ are clamped to this distribution producing an an XOR distribution (figure \ref{fig:learning_crn}D). Initially, the free chemical Boltzmann machine produces a uniform distribution (figure \ref{fig:learning_crn}E). Then, after tuning the potential species of both the visible and hidden units, it ultimately produces an XOR (figure \ref{fig:learning_crn}F). This result, that an autonomous CBM can learn an XOR distribution, is an indication of the computational power of this model because XORs can be glued together in order to represent increasingly complex distributions. Note in this example, the parameter $\alpha$ is used to scale the counts of the toggle switch so they can be represented by a binary variable. We describe the learning system simulated in \ref{fig:learning_crn} B in detail in SI \ref{SI:LearningCBM}.

This construction is quite general and can, in theory, be applied to any detailed balanced CRN, not just the chemical Boltzmann machine construction used in the simulation. Finally, although in this situation learning is active constantly, it is not hard to augment the construction so learning can be turned on and off, for example by setting $\epsilon=0$ or having the potential clamping reactions catalyzed by a species which can be controlled like a switch. We emphasize that this construction is a fully autonomous CRN capable of adapting diverse internal models to arbitrary environmental distributions. The fact that our entire system is self-contained with no external inputs will allow us to analyze rigorously analyze its thermodynamics.

\section{Thermodynamics of Learning}
This section provides energetic and thermodynamic costs for learning and inference with the potential clamping reactions. In the constructions used in this paper, learning and inference are fundamentally the same process. In both cases, we start with a dbCRN $\mathcal{C}_G$ with equilibrium distribution $\pi_G$ and, either via the potential clamping process or the learning process, end up with a new distribution $\pi_{G'}$. The final distribution $\pi_{G'}$ can be physically realized in a variety of ways: by changing the energies $G' = G + \Delta$; by equipping $\mathcal{C}_G$ with potential species so that $G' = G + \mu$; or by using the reactions (\ref{eq:learningCRN1}) to modulate the potentials species $\mathcal{P}$ in order to set a new equilibrium for the $\mathcal{S}$ species: $\pi_{G'}(s) = \pi^P(s)$. Importantly, in the first two scenarios described, $\pi_{G'}$ remains a dbCRN while in the last scenario $\pi_{G'}$ is out of equilibrium (at least until the potential clamping reactions are turned off). In related work, we provided idealized thermodynamic costs for the first two approaches assuming perfect recoverable modulation of the energies $\Delta$ and/or potentials $[P]$ \cite{poole2022detailed}. In this paper, we are only going to analyze the changes in the potential species $P$ due to the non-equilibrium potential clamping reactions. Although $P$ also changes due to its coupling with $S$ inside the detailed balanced vesicles, by construction this coupling will always be quasi-static which we have previously proved will require only reversible chemical work which is not dissipative \cite{poole2022detailed}.

In the following section we will first analyze the general dissipation for non-equilibrium steady-states induced by these reactions and  point out some trade-offs between accuracy, reversibility, and dissipation for this architecture. We will then apply these results towards understanding the learning CRN architecture.

\subsection{Thermodynamics of the Potential Clamping Reactions}
\label{sec:thermodynamics_of_clamping_crn}
\begin{figure}
    \centering
    \includegraphics[width = .95\textwidth]{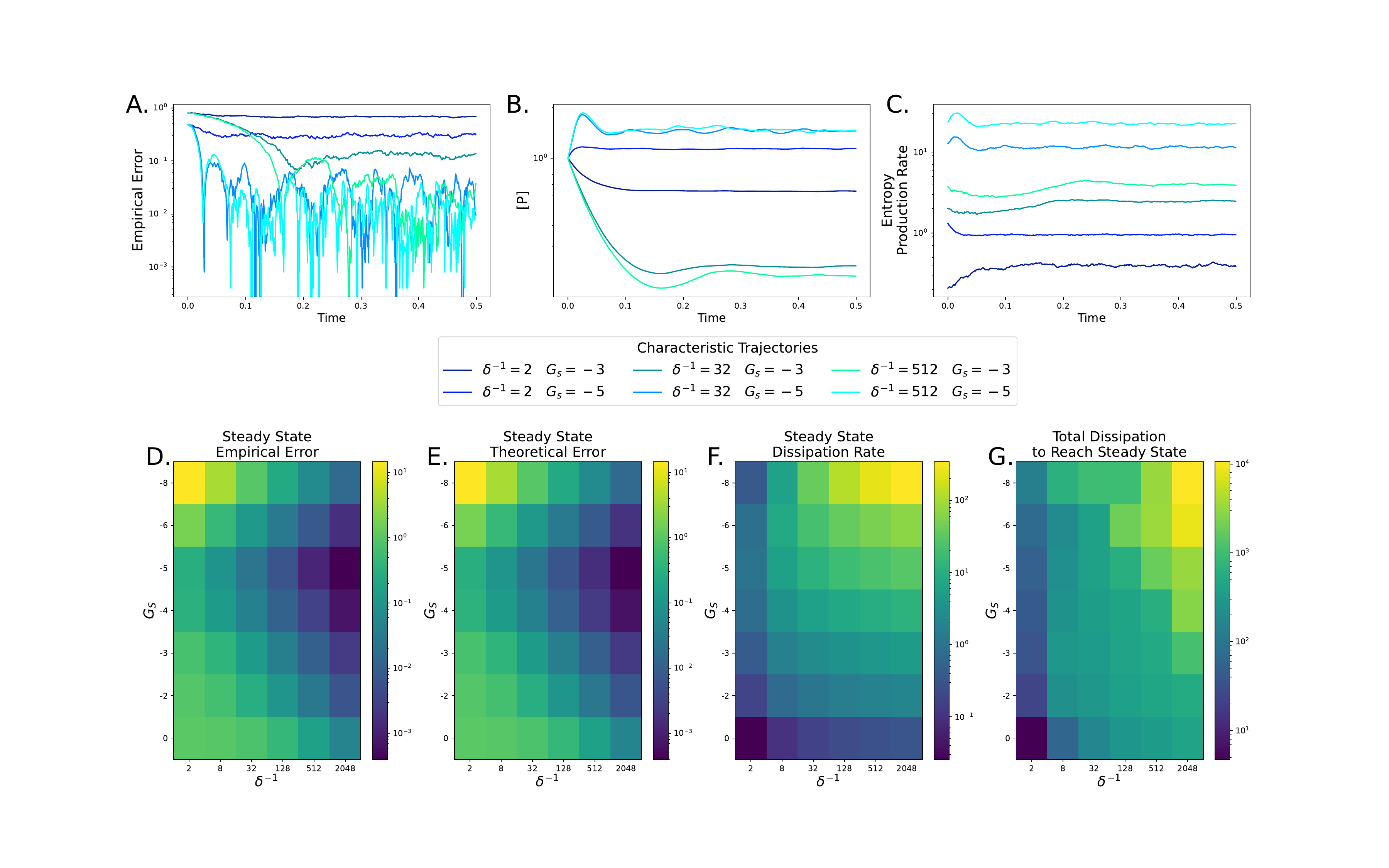}
    \caption{Error, entropy production rate, and total entropy production for the potential clamping CRN reactions applied to the dbCRN $\emptyset \xrightleftharpoons[]{} S$. A. The empirical error over the course of a simulation for various values of $\delta$ and $G_S$. B. The potential species concentration $[P]$ over the course of a simulation for various values of $\delta$ and $G_S$. C. The entropy rpoduction rate over the course of a simulation for various values of $\delta$ and $G_S$. D. Final empirical steady state error for various values of $\delta$ and $G_S$. E. Final theoretical error for various values of $\delta$ and $G_S$. F. Steady state dissipation rate for various values of $\delta$ and $G_S$. G. Total energy dissipated until convergence to steady state for various values of $\delta$ and $G_S$.}
    \label{fig:poisson_therm_sweep}
\end{figure}

To begin, we note the reactions (\ref{eq:learningCRN1}) are not directly amenable to thermodynamic treatment because they are irreversible and therefore infinitely dissipative. However, these equations can be rewritten reversibly and analyzed:
\begin{align}
    \label{eq:learningCRN_reversible1}
    P_i + S_i &\xrightleftharpoons[\delta]{\epsilon k_i^S} 2 P_i + S_i \\
    P_i + Q_i &\xrightleftharpoons[\delta]{\epsilon k_i^Q} Q_i,
    \label{eq:learningCRN_reversible2}
\end{align}
where $\delta$ is a small reverse rate constant. This CRN can be thought of as driven by a hidden infinite reservoir of fuel molecules~\cite{polettini2014irreversible}. To analyze the thermodynamics of the potential clamping reactions, we consider the fluxes through (\ref{eq:learningCRN_reversible1}), $J_i^S$, and (\ref{eq:learningCRN_reversible2}), $J_i^Q$ given by:
\begin{equation}
\label{eq:reaction_fluxes}
J_i^S = \epsilon [P_i] [S_i] - \delta [P_i]^2 [S_i] \quad \quad J_i^Q = \epsilon [P_i] [Q_i] - \delta [Q_i].
\end{equation}
Here we have assumed $k_i^S = k_i^Q = 1$ for simplicity. At steady-state, time derivative of $[P_i]$ must vanish. Because the coupling of $[P_i]$ with $S_i$ in the detailed balanced reactions is as quasi-equilibrium, this implies that the fluxes must either be 0 or cancel each other out:

\begin{align*}
    \frac{d[P_i]}{dt} = 0 \implies J_i^S - J_i^Q = 0 \quad \quad
    \textrm{Solution 1: } J_i^S = J_i^Q = 0 \quad \quad
    \textrm{Solution 2: } J_i^S = J_i^Q \neq 0.
\end{align*}
The first solution corresponds to an equilibrium solution when the driving potential goes to 0. In general, we do not expect this case to exhibit accurate learning. The second solution corresponds to a non-equilibrium steady state. Such a solution may exist provided that $\psi(q)$ has the same support as $\pi^P(s)$. A little algebraic manipulation reveals that at steady state the error between the mean of $S_i$ and $Q_i$ is given by:
\begin{align}
    \frac{\langle q_i \rangle_{\psi} - \langle s_i \rangle_{\pi^P}}{\langle q_i \rangle_{\psi}} &= 1 - \frac{\langle s_i \rangle_{\pi^P}}{\langle q_i \rangle_{\psi}} = 1 -\frac{(\epsilon [P_i] - \delta)}{[P_i](\epsilon - \delta [P_i])} = \frac{\delta (1 - [P_i]^2)}{[P_i](\epsilon - \delta [P_i])}
\end{align}
Notice that when $\delta \to 0$, the error also goes to 0. For non-zero values of $\delta$, the error will also depend non-linearly on the final concentration $[P_i]$. This dependence is illustrated in figure \ref{fig:poisson_therm_sweep}. The key insight from this figure is that the potential learning CRN can be only weakly driven and still work well, provided that the target mean $\langle q \rangle_\psi$ of the species $\mathcal{Q}$ is not to far from the unclamped mean $\langle s \rangle_\pi$ of the species $\mathcal{S}$. However, as larger clamping potentials need to be applied, the reverse rate constant $\delta$ must become small to keep the error low.

The reaction fluxes (\ref{eq:reaction_fluxes}) also allow us to calculate the entropy production rate from the potential clamping reactions using the thermodynamic results developed in \cite{rao2016nonequilibrium}:
\begin{align}
    \label{eq:clamping_dissipation}
    T \frac{d \mathbb{S}_i}{dt} = RT(J_i^S \log \frac{\epsilon}{\delta [P_i]} + J_i^Q \log \frac{\epsilon[P_i]}{\delta}).
\end{align}
Here, $R$ is the gas constant. At an equilibrium steady state, this expression simplifies to 0. By construction, $P$ vary due to the potentially clamping reactions quasi-statically so no energy is dissipated through the detailed balanced reactions \cite{poole2022detailed}. Therefore, the potential clamping reactions are the only parts of the system that produce entropy.

At a non-equilibrium steady state with $J_{ss} = J_i^S = J_i^Q$ the steady state entropy production is:
\begin{align}
    \frac{d \mathbb{S}_i^{ss}}{dt} = 2RTJ_{ss} \log \frac{\epsilon}{\delta}.
\end{align}
We show the entropy production (\ref{eq:clamping_dissipation})  applied to a simple birth death process $\emptyset \xrightleftharpoons[]{} S + P$ clamped to a constant $Q$ in figure (\ref{fig:learning_crn_dissipation}). As $\delta$ decreases, the error decreases at the expense of the steady state entropy production increasing. Interestingly, the entropy rate over time can be both an increasing or decreasing function depending on the energy of the species being clamped.

\subsection{Thermodynamics of the Learning CRN}
\begin{figure}
    \centering
    \includegraphics[width = .95\textwidth]{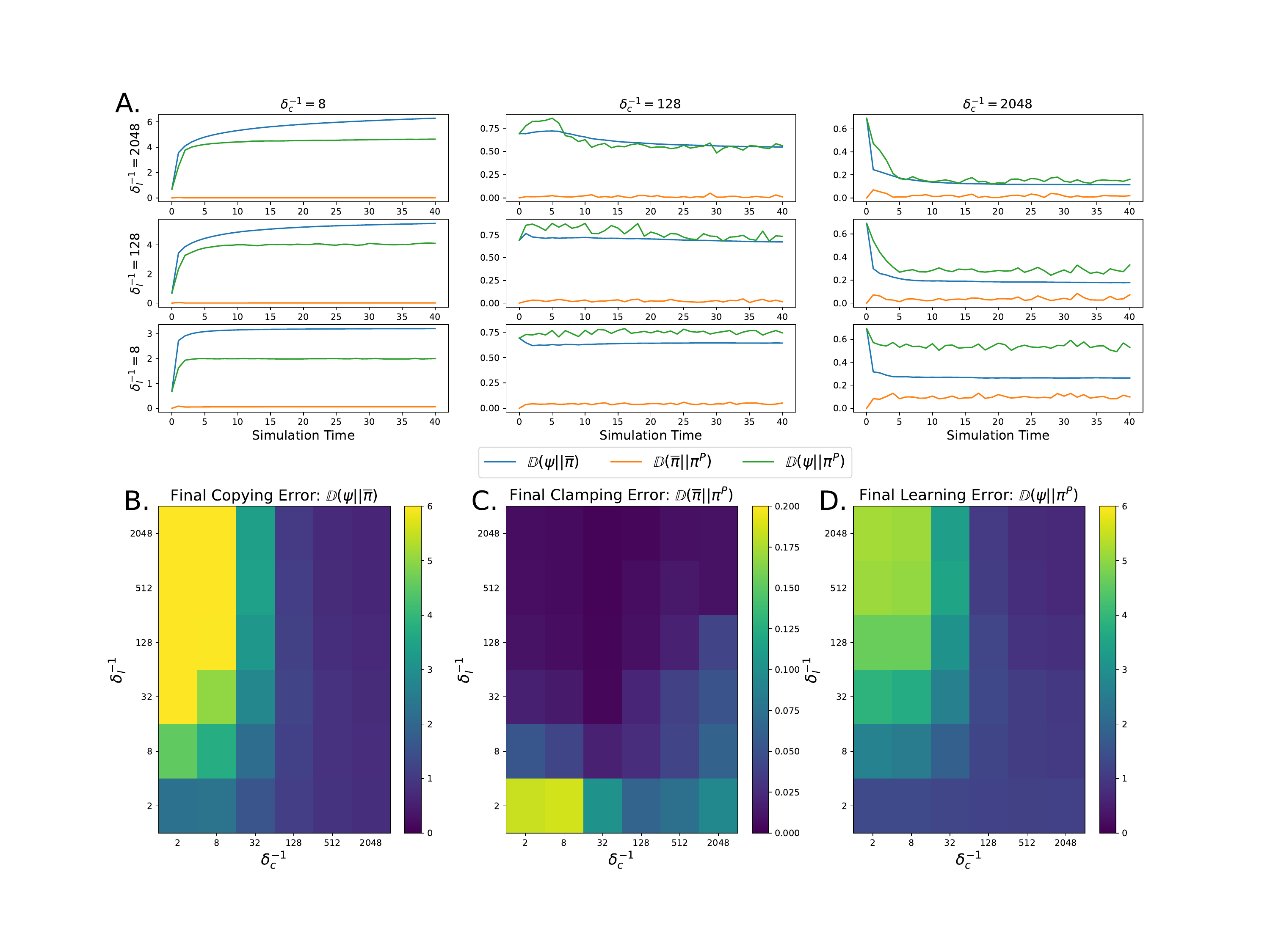}
    \caption{Thermodynamics of learning for the XOR dbCRN construction. A. The relative entropy between the clamped species and the environment (copying error), between the clamped species and free species (clamping error), and between the environment and free species (learning error) for various values of the reverse reaction rates $\delta_c$ and $\delta_l$ over the entire learning trajectory. B. The final copying error, defined as the relative entropy between the environment distribution and the clamped distribution as a function of the reverse rate constants. C. The final clamping error, defined as the relative entropy between the clamped distribution and the free distribution as a function of the reverse rate constants. D. The final learning error, defined as the relative entropy between the environment distribution and the free distribution as a function of the reverse rate constants.}
    \label{fig:learning_crn_dissipation}
\end{figure}

We conducted similar analysis on the XOR learning CRN model. In these simulations, the potential clamping reactions are separated into two sets each controlled by their own reverse rate constants. The first set are the reactions copying the environment to the clamped dbCRN, $\mathcal{T}^{\overline{\mathcal{P}}}_{\overline{\mathcal{S}}, \mathcal{Q}}$, with the reverse rate constant: $\delta_{c}$. The second set are the reactions coupling the clamped dbCRN to the free dbCRN, $\mathcal{T}^{\mathcal{P}}_{\mathcal{S}, \overline{\mathcal{S}}}$, with reverse rate constant $\delta_{l}$. By varying $\delta_{c}$ and $\delta_{l}$ independently while keeping all other rates of the simulation constant, we investigated the interplay between two non-equilibrium aspects of our construction: copying the environment into the system and learning internal parameters to model the environmental distribution. These results are illustrated in Figure (\ref{fig:learning_crn_dissipation}), which shows the copying error $\mathbb{D}(\psi || \overline{\pi})$ between the environment and the clamped dbCRN; the clamping error $\mathbb{D}(\overline{\pi} || \pi^P)$ between the clamped dbCRN and the free dbCRN; and the learning error $\mathbb{D}(\psi || \pi^P)$ between the environment and the free dbCRN. Where $\mathbb{D}$ is the relative entropy, $\psi$ is the environment distribution rescaled to $\{0, 1\}^2$, $\overline{\pi}$ is the clamped dbCRN distribution, and $\pi^P$ is the free dbCRN distribution.  In particular, both $\delta_l$ and $\delta_c$ need to be small in order for the autonomous learning CRN to have low relative entropy with the environment. Additionally, the copying error seems to define a lower limit to the learning error, suggesting that energy efficient learning must simultaneously optimize both of these processes. 

\section{Discussion}
We have presented a general chemical system capable of converting any potentiated dbCRN into an autonomous learning system. Underlying this construction are a set of potential clamping reactions that continuously modulate the chemical potentials of the species in the dbCRN. We have shown that these reactions implement a CRN version of gradient descent on the relative entropy which, intriguingly, also takes the form of a feedback control circuit. Unlike previous chemical implementations of learning algorithms \cite{hjelmfelt1992chemical,kim2004neural,mcgregor2012evolution,banda2013online,lakin2016supervised,poole2017chemical,anderson2021reaction,poole2022detailed}, our construction continuously and dynamically changes its parameters (the chemical potentials) in real time without external intervention. We further showed that the potential clamping reactions can be used to implement a continuous-time online version of the classic Boltzmann machine gradient descent learning algorithm, where one set of potential clamping reactions copies environmental variables into a clamped dbCRN and the second set of potential clamping reactions minimizes the relative entropy between the clamped dbCRN and a free dbCRN. This process allows for the free dbCRN to minimize its relative entropy with the environment utilizing hidden species to increase the complexity of the distributions it can represent. Finally, the concrete nature of our implementations allowed us to unambiguously analyze the thermodynamic costs of potential clamping and learning, adding to a growing body of work investigating the thermodynamics of machine learning algorithms.

One particularly novel aspect of our construction is its use of an ensemble of stochastic vesicles. We note that this choice may not be strictly necessary as a hybrid deterministic-stochastic CRN model with large separation of timescales can give similar behavior \cite{PooleThesis2021}. However, using an ensemble of CRN-vesicles has a number of potential applications. First, such a system could be implemented in a laboratory using droplet based technologies with potential species diffusing between droplets \cite{dupin2019signalling}. Second, the many-vesicle construction may also inspire novel machine-learning algorithms. Although our current \textit{in silico} implementation is reliant on conventional chemical reaction network simulators and therefore involves minimal parallelism, it may be possible to exploit the ensemble-of-vesicles interpretation of this model to develop highly parallel \textit{in silico} machine learning architectures based upon chemical reaction network models.

Finally, we wish to comment on the importance of our results to understanding diverse biological systems. 
Our construction suggests that any multi-compartment system which communicates between compartments via diffusive molecules could, potentially, be performing an algorithm similar to what we have described in this paper. Quorum sensing molecules such as AHLs in bacteria \cite{miller2001quorum} and morphogens in eukaryotic development \cite{rogers2011morphogen} are two biological examples where diffusive molecules may be acting as adaptive global potentials biasing groups of cells to perform specific functions. Additionally, as we mention above, our construction is not necessarily reliant on multiple vesicles and, instead, could be reliant on seperation of timescales with a quickly equilibrating dbCRN modulating itself via slowly changing potential species. The epigenetic code and corresponding chromatin structure may be an example of such a system. Transcription factor binding and covalent histone and DNA modifications can occur relatively quickly compared to the time it takes for transcription, translation, and maturation of new chromatin-modifying enzymes. In this sense, chromatin-modifying proteins could be analogous to potential species with the chromatin structure analogous to the dbCRN. Obviously biology is unlikely to exactly implement the learning constructions described in this paper, but we posit that understanding biological computation from a machine-learning inspired lens may prove fruitful.

\section{Acknowledgements:} We would like to thank E. Winfree and R.M. Murray of Caltech for helpful feedback on this project. W.P. was supported by the U.S. National Science Foundation. Additionally, the NSF in conjunction with the Indo-US Science and Technology Forum provided support for W.P. to spend a semester in Mumbai India working with M.G. which was instrumental to the success of this project.

\section{Methods:} All numerical simulations of chemical reaction networks were carried with the Bioscrape python package \cite{pandey2023fast} using a hybrid deterministic-stochastic approach. Tyically $N=10$ vesicles containing independent detailed balanced chemical reaction networks were simulated via the Gillespie algorithm. Simultaneously, bioscrape rules were used to numerically integrate the dynamics of the potential species every $dt$. This results in a coupled deterministic-stochastic hybrid simulation which is accurate for sufficiently small $dt$. A value of $dt=0.001$ was determined to be sufficiently accurate provided the smallest rate in the system $k_learn \geq 0.01$ by comparing increasingly smaller values of $dt$ and determining that the numerical output had sufficiently converged. Steady state distributions were found by simulating all systems for a long time until the distribution converged.

\section{Supplementary Information}

\subsection{Example of a potential clamped dbCRN with multiple steady-states:}
Consider the potentiated dbCRN:
\begin{align}
    S_0 \xrightleftharpoons[]{} P_1 + S_1 \quad \quad S_0 + S_2 \xrightleftharpoons[]{} S_0 + 2 S_2
\end{align}
with potential clamping reactions clamping $S_1$:
\begin{align}
\label{eq:pot_clamp_example}
    P_1 + Q_1 \to Q_1 \quad \quad P_1 +S_1 \to 2 P_1 + S_1
\end{align}
Notice that if the system begins in a state $(s_0, s_1, s_2, p_1, q_1) = (N_0, N_1, N_2, N_P, 1)$, then for all time the following conditions will hold: $s_0 + s_1 = N_0+N_1$, $s_2 \in [1, \infty]$, $p_0 \in [0, \infty)$, and $q_1=1$. Consider a specific trajectory where all of $S_0$ is converted to $S_1$ and $P_1$, and then all of $P_1$ is consumed by the first potential clamping reaction. Throughout this process, we allow $S_2$ to change to any value $N'_2$. This will bring the CRN to a state $(s_0, s_1, s_2, p_1, q_1) = (0, N_1+N_0, N'_2, 0, 1)$ at which point it will halt because there are no $S_0$ or $P_1$ species, so none of the reactions can fire. Importantly, this behavior can occur for any value $N'_2$ demonstrating the existence of multiple steady state distributions (which are all $\delta$ functions).

\subsection{Formal Derivation of the Quasi-equilibrium Vesicle Ensemble}
\label{si:qev_ensemble}

In this section we derive a deterministic non-mass action CRN from an ensemble of $N$ detailed balanced CRNs coupled via the potential clamping reactions by taking the limit of infinite vesicles in an infinite volume with defined concentration and limit of separation of timescales between the potential clamping reactions and the detailed balanced reactions. An $N-$ vesicle system can be represented as a single stochastic CRN $\mathcal{V}$ with unique species $\mathcal{S}^v$ in each vesicle and potential species $\mathcal{P}$ shared between vesicles:
\begin{align}
    \mathcal{V} = (\mathcal{S} \cup \mathcal{P}, \mathcal{R} \cup \mathcal{T}^\mathcal{P}_{\mathcal{S}, \mathcal{Q}}, k) \quad \textrm{with} \quad \mathcal{S} = \bigcup_{v=1}^{N} \mathcal{S}^v \quad \textrm{and} \quad \mathcal{R} = \bigcup_{v=1}^N \mathcal{R}^v.
\end{align}
Here, the species $\mathcal{S}^v$ react with each other via reactions $\mathcal{R}^v$ identically in each vesicle $v$ and the potential clamping reactions $\mathcal{T}^\mathcal{P}_{\mathcal{S}, \mathcal{Q}}$ include species $\mathcal{S}$ in all the vesicles. The dynamics of the entire system are given exactly using the master equation:
\begin{align}
    \label{si:split_cme}
    \frac{\textrm{d}\mathbb{P}(x, t)}{\textrm{d}t} &= 
    \epsilon_{slow} \left [ \sum_{r \in \mathcal{T}_{\mathcal{S}, \mathcal{Q}}^\mathcal{P}} \mathbb{P}(x - M^r, t) k_r \tilde{\rho}_r(x - M^r) - \mathbb{P}(x, t) k_r \tilde{\rho}_r(x) \right ]  \notag 
    \\ &+ 
    \epsilon_{fast} \left [ \sum_{r \in \mathcal{R}} \mathbb{P}(x - M^r, t) k_r \tilde{\rho}_r(x - M^r) - \mathbb{P}(x, t) k_r \tilde{\rho}_r(x) \right ].
\end{align}
Here, $x = (s^1....s^v, p)$ is a vector of all the species in each vesicle as well as the potential species and $\tilde{\rho}_r = \rho_r / (k_r \epsilon_r)$ is the propensity divided by the rate constant. The reactions have been split into two sets: the first summation is over the slow potential clamping reactions with a characteristic slow rate $\epsilon_r = \epsilon_{slow}$ assumed to be multiplied by all the rate constants $k_r$. Similarly, the second summation is over all the fast detailed balanced reactions occurring in the vesicles assumed to be multiplied by a characteristic fast rate $\epsilon_r = \epsilon_{fast}$.

We are interested in the limit of the dynamics as ${\epsilon_{fast}} \to \infty$. First we rewrite the chemical master equation (CME) in this limit:

\begin{align}
    &\lim_{\epsilon_{fast} \to \infty} \frac{1}{\epsilon_{fast}} \frac{\textrm{d}\mathbb{P}(x, t)}{\textrm{d}t} =  \notag \\
    &\lim_{\epsilon_{fast} \to \infty}
    \frac{\epsilon_{slow}}{\epsilon_{fast}} \left [ \sum_{r \in \mathcal{T}_{\mathcal{S}, \mathcal{Q}}^\mathcal{P}} \mathbb{P}(x - M^r, t) k_r \tilde{\rho}_r(x - M^r) - \mathbb{P}(x, t) k_r \tilde{\rho}_r(x) \right ]  \notag 
    \\ &\quad \quad + 
    \left [ \sum_{r \in \mathcal{R}} \mathbb{P}(x - M^r, t) k_r \tilde{\rho}_r(x - M^r) - \mathbb{P}(x, t) k_r \tilde{\rho}_r(x) \right ].
\end{align}
Taking the limit then results in the equation:
\begin{align}
    \label{si:vesicle_steady_state}
    0 &= \sum_{r \in \mathcal{R}} \mathbb{P}(x - M^r, t) k_r \tilde{\rho}_r(x - M^r) - \mathbb{P}(x, t) k_r \tilde{\rho}_r(x)
\end{align}
Notice that this equation is effectively asking for a steady-state solution to the fast reactions for all vesicles. By construction, these vesicles only share the potential species $p$. The following claim shows that these vesicles can in fact be treated independently.
\\\\
\textbf{Claim 1:} in the limit $p_i \gg s^v_i \, \forall \, i, \, \forall \, v$ in a potentiated CRN, the steady state distribution of a single vesicle $v$ is an equilibrium distribution:
\begin{align}
\label{si:single_vesicle_solution}
    \pi^P_v(s^v) = \pi_v(s^v \mid p) = \frac{1}{Z} e^{-\sum_i G_i s^v_i - G^P_i s^v_i - \log s^v_i! - s^v_i \log p_i}.
\end{align}
Here, $s_i^v$ is the count of species $i$ in vesicle $v$, $p_i$ is the (approximately constant) count of potential species $P_i$ and $G^P_i$ is the the energy of potential species $P_i$. 
\\
\textbf{Proof:} See supplemental section \textit{Proof of Energy Clamping and Potential Clamping Equivalence} in \cite{poole2022detailed}. 
\\\\
\textbf{Claim 2:} Equation (\ref{si:vesicle_steady_state}) has a solution given by:
\begin{align}
    \pi(x) = \pi(s, p) = \prod_{v=1}^N \pi^P_v(s^v).
\end{align}
Note that the dependence on $p$ is implied by the superscript $P$. This equation is important: the factored form of the distribution proves that all the vesicles are identically and independently distributed (i.i.d.). 
\textbf{Proof:} Equation (\ref{si:vesicle_steady_state}) can be rewritten:
\begin{align}
    \label{eq:quasi_equilibrium_condition}
    0 = \sum_{v=1}^N \sum_{r \in \mathcal{R}^v} \mathbb{P}(x - M^r, t) k_r \tilde{\rho}_r(x - M^r) - \mathbb{P}(x, t) k_r \tilde{\rho}_r(x).
\end{align}
The distribution $\mathbb{P}(x) = \pi_v(s^v)$ equates to 0 for each set of terms $r \in \mathcal{R}^v$ in the second summation by construction (it is the steady state of a single vesicle and there are no reactions occurring between species in different vesicles) completing the proof.
\\\\
This result for $\pi(s, p)$ in the limit $p_i \gg s_i^v$ is the quasi-equilibrium solution to the so-called \textit{fast manifold} and based upon singular perturbation theory can be inserted into the CME to understand the slow dynamics \cite{del2015biomolecular,plesa2023stochastic}. 

Next we rewrite the equation (\ref{si:split_cme}) in conditional form and take the time derivative:
\begin{align}
    \label{eq:conditional_derivative}
    \mathbb{P}(x, t) = \mathbb{P}(s, t \mid p) \mathbb{P}(p, t)
    \\
    \frac{\textrm{d}}{\textrm{d}t} \left[ \mathbb{P}(s, t \mid p) \mathbb{P}(p, t) \right] = \frac{\textrm{d} \mathbb{P}(s, t \mid p)}{\textrm{d}t}
    \mathbb{P}(p, t) + \mathbb{P}(s, t \mid p) \frac{\textrm{d} \mathbb{P}(p, t)}{\textrm{d}t} 
\end{align}
Based upon equation (\ref{eq:quasi_equilibrium_condition}) and the previous two claims, the time derivative of the fast reactions will be 0.
\begin{align}
\label{eq:quas_equilibrium_derivative}
\textrm{Quasi Equilibrium} \implies \mathbb{P}(s, t \mid p) = \pi^P(s) \implies \frac{\textrm{d} \mathbb{P}(s, t \mid p)}{\textrm{d}t} = \frac{\textrm{d} \pi^P(s)}{\textrm{d}t} = 0.
\end{align}
The slow reactions $r \in \mathcal{T}_{\mathcal{S}, \mathcal{Q}}^\mathcal{P}$, only vary the potential species ($S$ are catalytic). Combining the equations (\ref{si:split_cme}) , (\ref{eq:quasi_equilibrium_condition}), and (\ref{eq:quas_equilibrium_derivative}) results in:
\begin{align}
    &\frac{\textrm{d}}{\textrm{d}t} (\mathbb{P}(s, t \mid p)\mathbb{P}(p, t))  \notag \\
    &= 
    \frac{\textrm{d} \pi^P(s)}{\textrm{d}t}
    \mathbb{P}(p, t) + \pi^P(s) \frac{\textrm{d} \mathbb{P}(p, t)}{\textrm{d}t} 
    =
    0 + \pi^P(s) \frac{\textrm{d} \mathbb{P}(p, t)}{\textrm{d}t} 
    \\
    &=
     \epsilon_{slow} \pi^P(s) \left [ \sum_{r \in \mathcal{T}_{\mathcal{S}, \mathcal{Q}}^\mathcal{P}} \mathbb{P}(p - M^r, t) k_r \tilde{\rho}_r(s, p - M^r) - \mathbb{P}(p, t) k_r \tilde{\rho}_r(s, p) \right ] \\
    &=
    \epsilon_{slow} \left [
    \sum_{v = 1}^N \pi^P_v(s^v) \left ( \sum_{r \in \mathcal{T}_{\mathcal{S}^v, \mathcal{Q}}^\mathcal{P}} \mathbb{P}(p - M^r, t) k_r \tilde{\rho}_r(s^v, p - M^r) - \mathbb{P}(p, t) k_r \tilde{\rho}_r(s^v, p) \right )
    \right ].
\end{align}
In the second step, the fact that each vesicle $v$ has unique potential clamping reactions $\mathcal{T}_{\mathcal{S}^v, \mathcal{Q}}^\mathcal{P}$ was used to split the sum over vesicles $v$. Next we  take the limit of infinite vesicles in an infinite volume such that the vesicle concentration $N/V = \nu$:
\begin{align}
    &\lim_{\substack{N \to \infty \\ V \to \infty}} \frac{\textrm{d}}{\textrm{d}t} (\pi^P(s)\mathbb{P}(p, t))  \notag 
    \\
    &= \lim_{\substack{N \to \infty \\ V \to \infty}} \epsilon_{slow} \left [
    \sum_{v = 1}^N \pi^P_v(s^v) \left ( \sum_{r \in \mathcal{T}_{\mathcal{S}^v, \mathcal{Q}}^\mathcal{P}} \mathbb{P}(p - M^r, t) k_r \tilde{\rho}_r(s^v, p - M^r) - \mathbb{P}(p, t) k_r \tilde{\rho}_r(s^v, p) \right )
    \right ]
    \\
    &= \lim_{\substack{N \to \infty \\ V \to \infty}} \epsilon_{slow} N \sum_{r \in \mathcal{T}_{\mathcal{S}^v, \mathcal{Q}}^\mathcal{P}} \mathbb{P}(p - M^r, t) k_r\langle \tilde{\rho}_r(s^v, p - M^r) \rangle_{\pi^P} - \mathbb{P}(p, t) k_r\langle \tilde{\rho}_r(s^v, p) \rangle_{\pi^P}
    \\
    & =\lim_{\substack{N \to \infty \\ V \to \infty}} \epsilon_{slow} N \sum_i
    \mathbb{P}(p + 1_i, t) k_i^Q \langle \frac{q_i(p_i+1)}{V} \rangle_{\pi^P}
    + \mathbb{P}(p - 1_i, t) k_i^S \langle \frac{s_i(p_i-1)}{V} \rangle_{\pi^P}  \notag 
    \\ & \quad \quad \quad - \mathbb{P}(p, t) \left (k_i^Q \langle  \frac{p_i q_i}{V}\rangle_{\pi^P} + k_i^S \langle \frac{p_i s_i}{V} \rangle_{\pi^P} \right)
    \\
    &= \epsilon_{slow} \sum_i 
    \mathbb{P}(p + 1_i, t) k_i^Q \nu \langle q_i(p_i+1) \rangle_{\pi^P}
    + \mathbb{P}(p - 1_i, t) k_i^S \nu \langle s_i(p_i-1) \rangle_{\pi^P}  \notag 
    \\ & \quad \quad \quad- \mathbb{P}(p, t) \nu \left (k_i^Q \langle  p_i q_i\rangle_{\pi^P} + k_i^S \langle p_i s_i \rangle_{\pi^P} \right) 
    \label{si:limit_cme}
\end{align}
In the first step, the fact that each vesicle is i.i.d. is used to convert the sum into an expected value over the propensities which is valid in the $N\to \infty$ limit. In the second step, the potential clamping reactions have been explicitly written in the CME with $1_i$ denoting a unit vector containing the species $P_i$. In the final step, the $\nu = N/V$ is substituted to finish evaluating the limit.

By inspection, equation (\ref{si:limit_cme}) is equivalent to a chemical reaction network with non-mass-action dynamics:
\begin{align}
    &P_i + Q_i \xrightarrow{\rho_i^Q(q, p, \pi^P)} Q_i
    &&\rho_i^Q(q, p, \pi^P) = \epsilon_{slow}k_i^Q \nu \langle q_i, p_i \rangle_{\pi^P}
    \\
    &P_i + S_i \xrightarrow{\rho_i^S(s, p, \pi^P)} 2 P_i + Si
    &&\rho_i^S(s, p, \pi^P) = \epsilon_{slow} k_i^S p_i \nu \langle s_i \rangle_{\pi^P}.
\end{align}
Note that in the main text, we assumed the first reaction occurs in the bulk (not the vesicles) and removed the expected value from the propensity. However, this theory could also apply to clamping to a target species $Q_i$ inside of vesicles as we show more generally here. Finally, we comment that although this non-mass-action CRN could conceptually be analyzed with stochastic dynamics, the assumption that $p_i \gg s_i$ required in this derivation suggests that deterministic dynamics are appropriate due to the large copy number of $p_i$ required for this existing behavior.

\subsection{Connecting the Potential Clamping Reactions to Feedback Control}
\label{SI:feedback_control}
We will show that the potential clamping reactions (\ref{eq:potential_clamping_reactions}) are a form of integral feedback. In this context, integral feedback control simply means that a controlled state (or variable or species) will have a time derivative proportional to the error between that state and a reference (or target) signal.

Proof that the clamping CRN is a form of feedback control.

We will consider a single potentiated species $S_0$ part of a larger detailed balanced chemical reaction network with reactions $\mathcal{R}$. The potential species $P_0$ provides feedback on $S_0$ by modulating the chemical potential via the potential clamping reactions (\ref{eq:potential_clamping_reactions}). We consider the expected dynamics of $S_0$ under the quasi-equilibrium assumption and show these dynamics are proportional to the error: $[Q_0] - \langle s_0 \rangle_{\pi^P}$. We will assume $\nu = k_0^Q = k_0^S = 1$ so as not to have to rescale the error for simplicity. Note that this analysis does not make use of the vesicle ensemble limit and is valid even for a single vesicle. Based upon the definitions in section \ref{subsec:inference_with_dbcrns} and proven in \cite{poole2022detailed}, we can write the potentiated dbCRN distribution:
\begin{equation}
\pi^P(s) = \frac{1}{Z} e^{-(\sum_i G_i s_i + \log s_i !) - s_0 \log [P_0](t)}
\end{equation}
Taking the expecte value results in:
\begin{align}
\frac{d}{dt} \langle s_0 \rangle_{\pi^P} &= \frac{d}{dt} \sum_s s_0 \pi^P(s)
=
\frac{d}{dt} \sum_s s_0 \frac{1}{Z} e^{-(\sum_i G_i s_i + \log s_i !) - s_0 \log [P_0](t)} \\
&=
\sum_s s_0 \left [ (\frac{d}{dt}Z^{-1}) e^{-(\sum_i G_i s_i + \log s_i !) - s_0 \log [P_0](t)} + (Z^{-1} (\frac{d}{dt} e^{-(\sum_i G_i s_i + \log s_i !) - s_0 \log [P_0](t)})) \right ] \\
&= \epsilon \langle s_0\rangle_{\pi^P}^2 \left(\langle s_0 \rangle_{\pi^P} -  [Q_0] \right) - \epsilon \langle s_0^2 \rangle_{\pi^P} \left(\langle s_0 \rangle_{\pi^P} -  [Q_0] \right) \\
&=
\epsilon \left (\langle s_0^2 \rangle_{\pi^P} - \langle s_0\rangle_{\pi^P}^2 \right ) \left([Q_0]  - \langle s_0 \rangle_{\pi^P} \right).
\end{align}
Notice that this final term is proportional to the error: $[Q_0] - \langle s_0 \rangle_{\pi^P}$ times the variance $\textrm{Var}_{\pi^P}s_0 =  \langle s_0^2 \rangle_{\pi^P} - \langle s_0\rangle_{\pi^P}^2$. This demonstrates that the potential clamping reactions are a form of integral feedback control. In deriving the results above, each derivative was considered separately and evaluated using the dynamics $\frac{\textrm{d} [P_i]}{\textrm{d}t} =  \epsilon \left( [P_i] \langle s_i \rangle_{\pi^P} - [P_i][Q_i] \right)$. Specifically:
\begin{align}
\frac{d}{dt} e^{-(\sum_i G_i s_i + \log s_i !) - s_0 \log [P_0(t)]} &= e^{-(\sum_i G_i s_i + \log s_i !)-s_0 \log [P_0](t)} \frac{-s_0}{[P_0](t)} \frac{d [P_0](t)}{dt} = Z \pi^P(s) \frac{-s_0}{[P_0](t)} \frac{d [P_0](t)}{dt} \\
&=
-Z \pi^P(s) s_0 \epsilon \left(\langle s_0 \rangle_{\pi^P} -  [Q_0] \right).
\end{align}
Similarly,
\begin{align}
\frac{d}{dt}Z^{-1} &= - Z^{-2} \sum_s \frac{d}{dt} e^{-(\sum_i G_i s_i + \log s_i !) - s_0 \log [P_0](t)} = Z^{-1} \sum_s \pi^P(s) s_0 \epsilon \left(\langle s_0 \rangle_{\pi^P} -  [Q_0] \right) \\
&= Z^{-1} \epsilon \langle s_0\rangle_{\pi^P}\left(\langle s_0 \rangle_{\pi^P} -  [Q_0] \right).
\end{align}

\subsection{Extended Description of the Autonomous Learning CRN}
\label{SI:LearningCBM}
In this section, we describe the example of the learning system simulated in \ref{fig:learning_crn} B detail. This simulation consisted of $N=10$ identical vesicles each with their own internal stochastic CRNs. These consist of the free and clamped CRNs, represented by species $S$ and $\overline{S}$, respectively, which are both 3-node chemical Boltzmann machines with two visible species $A$ and $B$ and three hidden species implementing a hidden node $H$ and the weights connecting the hidden unit to the visible units $W_{HA}$ and $W_{HB}$. Although each vesicle has its own internal stochastic CRN, the potential species are shared between all vesicles. This entire model takes the form of one large hybrid CRN with four nominal timescales; $\epsilon_{fast}$, the environmental clamping rate; $\epsilon_{slow}$, the hidden unit clamping rate; $k_{db}$, the detailed balanced CRN nominal rate; and $k_{env}$, the environmental rate. These rates are separated into timescales: $k_{env} c_{ev} \approx k_{db} * c_{db} \gg \epsilon_{fast} * C_p \gg \epsilon_{slow} * C_p$. Here, $c_{ev}$, $c_{db}$ are the characteristic counts of the environment and dbCRNs, respectively. $C_p$ is the characteristic concentration of the potential species. Due to the fact that the concentrations of the potential species are potentially unbounded, this implies learning will fail if the potential concentrations begin becoming very high. Such a situation could be encountered when the underlying dbCRN is not capable of learning the environmental distribution (e.g. no steady state exists for the potential species).
The environment $\psi(q^V, q^H)$ is generated by a bistable toggle switch consisting of  two genes $i \in \{A, B\}$; each gene, $G_i$, produces a transcript $T_i$  which is translated into a repressor $Q_i$. These repressors bind cooperatively to genes of the opposite type to form the repressed complex $C_{ij}$. Finally, a small amount of leak is added even for the repressed genes to help tune switching times and the transcripts and repressors degrade at rate $\delta$. Only the repressors are visible, $\mathcal{Q}^V = \{Q_A, Q_B\}$. All other species are hidden: $\mathcal{Q}^H = \{G_A, G_B, T_A, T_B, C_{AB}, C_{BA}\}$. We model this process with the stochastic mass action reactions:
\begin{align*}
    &G_i \xrightarrow{k_{tx}} G_i + T_i \quad \quad 
    T_i \xrightarrow{k_{tl}} T_i + Q_i \quad \quad
    G_i + 2Q_j \xrightleftharpoons[k_u]{k_b} C_{ij}
    \\
    &C_{ij} \xrightarrow{k_{leak}} C_{ij} + T_i \quad \quad
    T_i \xrightarrow{\delta} \emptyset \quad \quad 
    Q_i \xrightarrow{\delta} \emptyset \quad\quad (i \neq j)
\end{align*}
Here $i, j \in \{A, B\}$ and the rate constants $k_{tx}$, $k_{tl}$, $k_u$, $k_b$, $\delta \geq  k_{env}$. Next, the clamped potentiated detailed balanced reactions produce the distribution $\overline{\pi}^{\overline{P}, P}(s^V, s^H \mid \langle s^V \rangle = q^V) \psi(q^V)$. Internally, it is a 3-node ECBM with potentiated detailed balanced reactions:
\begin{align*}
\mathcal{R}^{\overline{S}}_{\overline{\mathcal{P}}, \mathcal{P}} = \{ \\
    \overline{S}^0_A + \overline{S}_H^0 + \overline{P}^0_{A}+P_A^0 &\xrightleftharpoons{}{}  \overline{S}^1_A+ \overline{S}_H^0 + \overline{P}^1_{A} + P_A^1
    \\
    \overline{S}^0_A + \overline{S}_H^1 + \overline{S}_{W_{HA}}^0 + \overline{P}^0_{A} + P_A^0 + \overline{P}_{W_{HA}}^0  &\xrightleftharpoons{}{}  \overline{S}^1_A+ \overline{S}_H^1 + \overline{S}_{W_{HA}}^1 + \overline{P}^1_{A} +P_A^1 + P_{W_{HA}}^1
    \\
    \overline{S}^0_B + \overline{S}_H^0 + \overline{P}^0_{B} + P_B^0 &\xrightleftharpoons{}{}  \overline{S}^1_{B}+ \overline{S}_H^0 + \overline{P}^1_{B} + P_B^1
    \\
    \overline{S}^0_B + \overline{S}_H^1 + \overline{S}_{W_{HB}}^0 + \overline{P}^0_{B}  + P_B^0 + P_{W_{HB}}^0  &\xrightleftharpoons{}{}  \overline{S}^1_B+ \overline{S}_H^1 + \overline{S}_{W_{HB}}^1 + P^1_{B} + P_B^1 + P_{W_{HB}}^1\\
    \overline{S}_H^0 + \overline{S}_A^0 + \overline{S}_B^0 +P_H^0 &\xrightleftharpoons{}{} \overline{S}_H^1 + \overline{S}_A^0 + \overline{S}_B^0 + P_H^1 \\
    \overline{S}_H^0 + \overline{S}_A^1 + \overline{S}_B^0 + \overline{S}_{W_{HA}}^0 + P_H^0 + P_{W_{HA}}^0 &\xrightleftharpoons{}{} \overline{S}_H^1 + \overline{S}_A^1 + \overline{S}_B^0 + \overline{S}_{W_{HA}}^1 + P_H^1 + P_{W_{HA}}^1\\
    \overline{S}_H^0 + \overline{S}_A^0 + \overline{S}_B^1 + \overline{S}_{W_{HB}}^0 + P_H^0 + P_{W_{HB}}^0 &\xrightleftharpoons{}{} \overline{S}_H^1 + \overline{S}_A^0 + \overline{S}_B^1 + \overline{S}_{W_{HB}}^1 + P_H^1 + P_{W_{HB}}^1\\
    \overline{S}_H^0 + \overline{S}_A^1 + \overline{S}_B^1 + \overline{S}_{W_{HA}}^0 + \overline{S}_{W_{HB}}^0 + P_{W_{HA}}^0 + P_H^0 + P_{W_{HB}}^0 &\xrightleftharpoons{}{} \overline{S}_H^1 + \overline{S}_A^1 + \overline{S}_B^1 + \overline{S}_{W_{HA}}^1 + \overline{S}_{W_{HB}}^1 + P_H^1 + P_{W_{HA}}^1 + P_{W_{HB}}^1
    \\ \}.
\end{align*}
Here, all the rate constants are detailed balanced and are scaled by $k_{db}$. The potential species $\overline{P}^1_A$ and $\overline{P}^1_B$ are used to clamp the visible species $\overline{S}_A$ and $\overline{S}_B$ to the visible environmental species $Q_A$ and $Q_B$ by the non-detailed balanced potential clamping reactions: 
\begin{align*}
\mathcal{T}^{\overline{\mathcal{P}}}_{\overline{\mathcal{S}}, \mathcal{Q}} = \{ \\
    &\overline{P}_{A}^1 + \overline{S}_A^1 
    \xrightarrow[]{\epsilon_{fast} k^{\overline{S}_A}}
    2 \overline{P}_{A}^1 + \overline{S}_A^1 
    \quad \quad 
    &&\overline{P}_{A}^1 + Q_A \xrightarrow[]{\epsilon_{fast} k^Q_A} Q_A 
    \\
    &\overline{P}_{B}^1 + \overline{S}_B^1 
    \xrightarrow[]{\epsilon_{fast} 
    k^{\overline{S}}_B} 2 \overline{P}_{B}^1 + \overline{S}_B^1 
    \quad \quad 
    &&\overline{P}_{B}^1 + Q_B \xrightarrow[]{\epsilon_{fast} k^Q_B} Q_B 
    \\ \}
\end{align*}

The rates of the above reactions are set to rescale the means of the repressors: $\frac{k_A^{\overline{S}}}{k_A^Q} = \frac{k_B^{\overline{S}}}{k_B^Q} = \frac{k_{tx}k_{tl}}{\delta^2}$.
Next, we describe the free potentiated detailed balanced reactions which produce the distribution $\pi^P(s^v, s^h)$. These reactions are near duplicates of the clamped potentiated detailed balanced reactions; the former lacks the potential species $\overline{P}^1_A$ and $\overline{P}^1_B$. Specifically, these reactions model a 3-node ECBM:

\begin{align*}
\mathcal{R}_\mathcal{P}^\mathcal{S} = \{ \\
    S^0_A + S_H^0 &\xrightleftharpoons{}{}  S^1_A+ S_H^0 + P^1_A
    \\
    S^0_A + S_H^1 + S_{W_{HA}}^0 + P_{W_{HA}}^0  &\xrightleftharpoons{}{}  S^1_A+ S_H^1 + S_{W_{HA}}^1 + P^1_A + P_{W_{HA}}^1
    \\
    S^0_B + S_H^0 &\xrightleftharpoons{}{}  S^1_B+ S_H^0 + P^1_B
    \\
    S^0_B + S_H^1 + S_{W_{HB}}^0 + P_{W_{HB}}^0  &\xrightleftharpoons{}{}  S^1_B+ S_H^1 + S_{W_{HB}}^1 + P^1_B + P_{W_{HB}}^1 \\
    S_H^0 + S_A^0 + S_B^0 +P_H^0 &\xrightleftharpoons{}{} S_H^1 + S_A^0 + S_B^0 + P_H^1 \\
    S_H^0 + S_A^1 + S_B^0 + S_{W_{HA}}^0 + P_H^0 + P_{W_{HA}}^0 &\xrightleftharpoons{}{} S_H^1 + S_A^1 + S_B^0 + S_{W_{HA}}^1 + P_H^1 + P_{W_{HA}}^1\\
    S_H^0 + S_A^0 + S_B^1 + S_{W_{HB}}^0 + P_H^0 + P_{W_{HB}}^0 &\xrightleftharpoons{}{} S_H^1 + S_A^0 + S_B^1 + S_{W_{HB}}^1 + P_H^1 + P_{W_{HB}}^1\\
    S_H^0 + S_A^1 + S_B^1 + S_{W_{HA}}^0 + S_{W_{HA}}^0 + P_{W_{HB}}^0 + P_H^0 + P_{W_{HB}}^0 &\xrightleftharpoons{}{} S_H^1 + S_A^1 + S_B^1 + S_{W_{HA}}^1 + S_{W_{HB}}^1 + P_H^1 + P_{W_{HA}}^1 + P_{W_{HB}}^1.
    \\ \}
\end{align*}
The free species $\mathcal{S}$ species are coupled to clamped $\overline{\mathcal{S}}$ species via the potential species $\mathcal{P}$ which are modulated by the potential clamping reactions:
\begin{align*}
\mathcal{T}^{\mathcal{P}}_{\mathcal{S}, \overline{\mathcal{S}}} = \{ \\
    &P_{A}^0 + S_A^0 \xrightarrow[]{\epsilon_{slow}} 2 P_{A}^0 + S_A^0 \quad \quad &&P_{A}^0 + \overline{S}_A^0 \xrightarrow[]{\epsilon_{slow}} \overline{S}_A^0 \\
    &P_{A}^1 + S_A^1 \xrightarrow[]{\epsilon_{slow}} 2 P_{A}^1 + S_A^1 \quad \quad &&P_{A}^1 + \overline{S}_A^1 \xrightarrow[]{\epsilon_{slow}} \overline{S}_A^1 \\
    &P_{B}^0 + S_B^0 \xrightarrow[]{\epsilon_{slow}} 2 P_{B}^0 + S_B^0 \quad \quad &&P_{B}^0 + \overline{S}_B^0 \xrightarrow[]{\epsilon_{slow}} \overline{S}_B^0 \\
    &P_{B}^1 + S_B^1 \xrightarrow[]{\epsilon_{slow}} 2 P_{B}^1 + S_B^1 \quad \quad &&P_{B}^1 + \overline{S}_B^1 \xrightarrow[]{\epsilon_{slow}} \overline{S}_B^1 \\
    &P_{H}^0 + S_H^0 \xrightarrow[]{\epsilon_{slow}} 2 P_{H}^0 + S_H^0 \quad \quad &&P_{H}^0 + \overline{S}_H^0 \xrightarrow[]{\epsilon_{slow}} \overline{S}_H^0 \\
    &P_{H}^1 + S_H^1 \xrightarrow[]{\epsilon_{slow}} 2 P_{H}^1 + S_H^1 \quad \quad &&P_{H}^1 + \overline{S}_H^1 \xrightarrow[]{\epsilon_{slow}} \overline{S}_H^1 \\
    &P_{W_{HA}}^0 + S_{W_{HA}}^0 \xrightarrow[]{\epsilon_{slow}} 2 P_{W_{HA}}^0 + S_{W_{HA}}^0 \quad \quad &&P_{W_{HA}}^0 + \overline{S}_{W_{HA}}^0 \xrightarrow[]{\epsilon_{slow}} \overline{S}_{W_{HA}}^0 \\
    &P_{W_{HA}}^1 + S_{W_{HA}}^1 \xrightarrow[]{\epsilon_{slow}} 2 P_{W_{HA}}^1 + S_{W_{HA}}^1 \quad \quad &&P_{W_{HA}}^1 + \overline{S}_{W_{HA}}^1 \xrightarrow[]{\epsilon_{slow}} \overline{S}_{W_{HA}}^1 \\
    &P_{W_{HB}}^0 + S_{W_{HB}}^0 \xrightarrow[]{\epsilon_{slow}} 2 P_{W_{HB}}^0 + S_{W_{HB}}^0 \quad \quad &&P_{W_{HB}}^0 + \overline{S}_{W_{HB}}^0 \xrightarrow[]{\epsilon_{slow}} \overline{S}_{W_{HB}}^0 \\
    &P_{W_{HB}}^1 + S_{W_{HB}}^1 \xrightarrow[]{\epsilon_{slow}} 2 P_{W_{HB}}^1 + S_{W_{HB}}^1 \quad \quad &&P_{W_{HB}}^1 + \overline{S}_{W_{HB}}^1 \xrightarrow[]{\epsilon_{slow}} \overline{S}_{W_{HB}}^1.
    \\ \}
\end{align*}
Notice that the visible clamped species $\overline{S}_A$ and $\overline{S}_B$ have their own potentials in $\overline{\mathcal{P}}$ as well as sharing potentials species in $\mathcal{P}$ with $S_A$ and $S_B$. However the hidden clamped species $\overline{S}_H$, $\overline{S}_{W_{HA}}$ and $\overline{S}_{W_{HB}}$ only share potentials $\mathcal{P}$ with the corresponding free species $S_H$, $S_{W_{HA}}$ and $S_{W_{HB}}$. This is reminiscent of the way the clamped units in a Boltzmann machine use the same energies as the free units during training.

\bibliography{bibliography}

\end{document}